\documentclass[floats,floatfix,showpacs,preprintnumbers,amssymb,prd,twocolumn,superscriptaddress,nofootinbib,nolongbibliography,reprint]{revtex4-1}

\usepackage{amssymb,amsmath,verbatim,mathtools,needspace,enumitem,etoolbox,graphicx,physics,microtype,afterpage,xspace,tabularx,lmodern,multirow,hhline,boldline, makecell, booktabs, subfigure}
\usepackage{gensymb}
\usepackage{appendix}
\usepackage{tensor}
\usepackage{array}
\usepackage{siunitx}

\makeatletter
\def\l@subsubsection#1#2{}
\def\l@subsubsubsection#1#2{}
\makeatother

\usepackage{comment}
\usepackage{mathrsfs}
\usepackage[usenames]{color}
\usepackage{epstopdf}
\usepackage{aas_macros}
\usepackage{bm}
\usepackage{dcolumn}
\usepackage{latexsym}
\usepackage{rotating}
\usepackage{longtable}

\usepackage{enumerate}
\usepackage{tensor,multirow}
\usepackage{url}
\usepackage{float}
\usepackage{array}
\newcolumntype{C}[1]{>{\centering\let\newline\\\arraybackslash\hspace{0pt}}m{#1}}

\usepackage[dvipsnames]{xcolor}
\usepackage[unicode]{hyperref}
\hypersetup{colorlinks=true, citecolor=MidnightBlue,
            linkcolor=MidnightBlue, urlcolor=MidnightBlue, linktocpage=true}
            
\usepackage{tikz}

\newcommand{\be}{\begin{equation}}
\newcommand{\eeq}{\end{equation}}
\newcommand{\ba}{\begin{align}}
\newcommand{\ea}{\end{align}}

\newcommand{\bcr}{b_{\rm{cr}}}
\newcommand{\bgr}{b_{\rm{gr}}}
\newcommand{\bmax}{b_{\rm{max}}}
\newcommand{\bmin}{b_{\rm{min}}}
\newcommand{\pd}{\partial}

\newcommand{\jhu}{William H.\ Miller III Department of Physics and Astronomy, Johns Hopkins University, \\ 3400 N. Charles Street, Baltimore, Maryland, 21218, USA}

\newcommand{\usd}{Department of Physics, University of South Dakota, Vermillion, SD 57069, USA}

\newcommand{\uvaphys}{Department of Physics, University of Virginia, P.O. Box 400714, Charlottesville, Virginia 22904-7414, USA}

\newcommand{\uvaastro}{Department of Astronomy, University of Virginia, 530 McCormick Road, Charlottesville, VA 22904, USA}

\newcommand{\iacs}{School of Physical Sciences, Indian Association for the Cultivation of Science, Kolkata 700032, India}

\newcommand{\GSSI}{Gran Sasso Science Institute (GSSI), I-67100 L’Aquila, Italy}
\newcommand{\GranSasso}{INFN, Laboratori Nazionali del Gran Sasso, I-67100 Assergi, Italy}

\begin{document}

\title{Relativistic dynamical friction and accretion for Kerr black holes}

\author{Nicholas Speeney}
\email{dst2bt@virginia.edu}
\affiliation{\uvaphys}
\affiliation{\uvaastro}

\author{Soumodeep Mitra}
\email{soumodeep.mitra@coyotes.usd.edu}
\affiliation{\usd}

\author{Emanuele Berti}
\email{berti@jhu.edu}
\affiliation{\jhu}

\author{Andrea Maselli}
\email{andrea.maselli@gssi.it}
\affiliation{\GSSI}
\affiliation{\GranSasso}

\author{Sumanta Chakraborty}
\email{tpsc@iacs.res.in}
\affiliation{\iacs}

\begin{abstract}
Extreme-mass-ratio inspirals (EMRIs) observed by LISA will probe the environments of massive black holes (BHs), including the dark matter (DM) overdensities expected to form around them. We compute the relativistic dynamical friction and accretion experienced by a non-spinning compact object on circular equatorial orbits around a Kerr BH embedded in a collisionless DM distribution, extending previous results in Schwarzschild. The DM is modeled as a spike grown adiabatically from an initial Hernquist profile, and the momentum transfer is obtained by integrating relativistic scattering and capture cross sections over the full, generically anisotropic phase space seen by the secondary. Far from the primary the friction tracks the local DM density, which increases with spin. Near the innermost stable circular orbit of rapidly spinning BHs ($a\gtrsim 0.37$), however, frame dragging forces bound DM particles to co-rotate, so that the secondary is overtaken by the surrounding medium: the friction weakens and eventually reverses sign where the DM density peaks. We show that a similar reversal can be mimicked around a non-rotating BH if the initial halo carries large net angular momentum, although this requires anisotropies beyond those found in cosmological simulations. We also show that commonly used Chandrasekhar-like prescriptions, which depend only on the local density, can overestimate the friction force. Our results demonstrate that the full phase-space structure of the environment must be included when modeling DM effects on EMRIs.
\end{abstract}

\maketitle
\section{Introduction}\label{sec:intro}
Understanding the nature of dark matter (DM) is one of the most pressing questions in modern science. Several lines of evidence (including observations of galaxy rotation curves, lensing measurements, and anisotropies in the cosmic microwave background) point to the existence of DM, and further show that it constitutes $\sim27\%$ of the energy density of the Universe, $\sim5\%$ of it being ``ordinary'' baryonic matter, and $\sim68\%$ dark energy~\cite{Freese:2008cz,Planck:2018vyg}. One of the most widely successful models for the DM which explains most of the observed phenomena is the ``cold dark matter'' (CDM) paradigm, in which the DM is made up of noninteracting, 
collisionless particles that interact with baryonic matter (and with other DM particles) only gravitationally.

The $\Lambda$CDM model of cosmology is very successful, but it faces some difficulties at sub-galactic scales~\cite{Bullock:2017xww}. For example, numerical simulations of galactic CDM halos have generally found results which do not agree with observed DM distributions (the ``cusp-core'' problem~\cite{Flores:1994gz,deBlok:2009sp}), and the standard $\Lambda$CDM model cannot explain the low abundance of dwarf galaxies in the local group (the ``missing satellites'' problem~\cite{Klypin:1999uc,Moore:1999nt}). These issues have led to a plethora of DM models which seek to remedy these problems, mostly originating from the particle physics community~\cite{Bertone:2004pz}. Despite large experimental efforts, no such DM particle has been detected to date, and the physical character of the DM remains elusive~\cite{Cirelli:2024ssz}.

At present, the best tests we can perform come from cosmological and astrophysical measurements of the gravitational effect of DM on luminous matter. However these probes are limited by angular resolution, which limits our ability to probe DM near galactic centers. Numerical simulations also face difficulties in this region, as their galactic-scale halo evolution cannot resolve the comparatively small region near the galactic center. To this end, one of the most promising avenues to explore both the nature of DM and its distribution in the centers of galaxies is through gravitational wave (GW) astronomy. 
The Laser Interferometer Space Antenna (LISA)~\cite{LISA:2017pwj,LISA:2022yao,LISA:2024hlh} will allow for the detection of low-frequency GW signals originating from a host of currently inaccessible sources. One such source consists of the extreme-mass-ratio inspiral (EMRI) of stellar-mass compact bodies into massive black holes (BHs). These systems are expected to form and evolve near galactic centers, and hence are powerful probes of the astrophysical environments there. EMRI systems may undergo $\sim10^5$ cycles within the LISA band throughout their lifetime, and small orbital deviations from beyond-GR effects, dynamical encounters, or interactions with the local environment may leave detectable imprints in the GW signals they produce. In particular, environmental effects in EMRI systems have become an area of great interest in recent years, garnering much attention from the GW community. Recent advances in precision vacuum and non-vacuum EMRI modeling have begun to pave the way for precision science with LISA.

Treatments of environmental effects in EMRIs have developed along two distinct but complementary lines. Perturbative methods have been employed in a variety of environmental scenarios, including dense anisotropic fluid environments~\cite{Cardoso:2021wlq,Cardoso:2022whc,Figueiredo:2023gas,Gliorio:2025cbh,Speeney:2024mas}, scalar fields~\cite{Xu:2026aic,Xu:2026cky,Brito:2023pyl,Duque:2023seg,Li:2025ffh}, and diffuse media admitting a two-parameter expansion in the mass ratio and environmental ``smallness''~\cite{Dyson:2025dlj,Datta:2025ruh,Rahman:2025mip,Dyson:2026ddd,Li:2025ffh,Li:2026foy}. These frameworks have the advantage of being fully relativistic and self-consistent, which is needed for precision waveform modeling. A disadvantage of such treatments is their complexity, which makes it difficult to treat non-idealized environmental scenarios and generic orbits within perturbative schemes. A detailed understanding of the environment's evolution and response to the secondary body's motion is also hard to incorporate. Alternatively, non-perturbative treatments adopting a ``zoomed-in'' approach have been developed and employed to treat similar problems~\cite{Vicente:2025gsg,Karydas:2025bkj,Karydas:2025gqp}. These approaches relate local accelerations to the secondary's orbital motion, in analogy with the gravitational self-force program~\cite{Barack:2009ux,Barack:2018yvs,Pound:2021qin,Poisson:2011nh}. Local interactions with the secondary and the environment naturally admit a phase space description, in which scattering and accretion of particles is encoded in the particle distribution function. It is also possible to consider back-reaction onto the environment~\cite{Kavanagh:2024lgq,Kavanagh:2020cfn,Coogan:2021uqv,Nichols:2023ufs,Wade:2025rkk,Wade:2026xvm}, which is hard to capture within perturbative schemes (but environmental depletion due to the secondary's motion has been incorporated in scalar field backgrounds~\cite{Li:2025ffh,Dyson:2025dlj}).

In this paper we study an EMRI evolving within a collisionless DM distribution in the ``zoomed-in'' approach. As the secondary orbits, particles in the surrounding medium encounter the orbiting body and may scatter or be accreted. These processes transfer momentum to the secondary, inducing changes to its orbital parameters alongside GW radiation-reaction. The scattering, or ``dynamical friction'' (DF), was first studied in the Newtonian regime by Chandrasekhar in the context of a body undergoing linear motion through an isotropic distribution of stars~\cite{1943ApJ....97..255C}. The body causes the formation of a higher density wake behind it, which in turn induces a net gravitational drag force acting against its motion. For bodies moving relativistically, treatments of DF in gaseous media~\cite{Barausse:2007ph,Katz:2019rgf}, scalar and bosonic fields~\cite{Traykova:2023qyv,Annulli:2020ilw,Vicente:2022ivh,Mitra:2026gar}, and collisionless particle distributions~\cite{Vicente:2025gsg} have been developed. These various treatments have been used to study the impact of environments on BH binary dynamics and their GW signals~\cite{Speeney:2022ryg,Mitra:2023sny,Mitra:2025tag,Karydas:2025bkj,Karydas:2025gqp,Duque:2023seg,Xiong:2026fqn}, although a treatment which is fully relativistic, incorporates the environmental back-reaction and evolution, and is valid in the full Kerr geometry is still lacking. In this work, we address the latter of these issues by computing the DF and accretion due to a collisionless DM halo in the Kerr background. We adopt the model of DF first derived in Ref.~\cite{Vicente:2025gsg}, and take the particle distribution to be the adiabatically grown DM spike profiles of Ref.~\cite{Ferrer:2017xwm}. We explicitly demonstrate the impact of the central BH spin on the density distribution and the allowed DM particle phase space, and connect these properties to features of the DF force. In the Kerr background, we find a sign reversal of the friction force for sufficiently high central BH spins, in line with recent work~\cite{Xiong:2026fqn}. We show that such reversals in the friction may be mimicked in non-rotating BH backgrounds by an initially spinning DM halo, suggesting that anisotropies in the chosen DM distribution function may significantly impact EMRI dynamics for commonly used adiabatic growth models of relativistic DM halos. 

The paper is structured as follows. In Sec.~\ref{sec:DF_acc} we turn to the problem of DF and accretion, and detail how a secondary BH in an EMRI interacts with a collisionless DM environment. In Sec.~\ref{sec:adiabatic_growth}, we review adiabatic growth models of a DM halo in the Kerr background, first developed in~\cite{Ferrer:2017xwm}. In Sec.~\ref{sec:results} we apply these results to EMRI systems evolving in DM halos, and show how frictional effects are impacted by rotating backgrounds and different choices of the distribution function. In Sec.~\ref{sec:conclusions} we present our conclusions and directions for future work. Throughout this paper we adopt geometrical units $(G=c=1)$. Greek indices run over spacetime coordinates, and Latin indices over spatial coordinates. Unless otherwise specified, we adopt dimensionless units and normalize all quantities to the central BH mass $M$.

\section{Dynamical Friction and Accretion in Kerr}\label{sec:DF_acc}

The formalism to compute relativistic friction and accretion for an EMRI on a circular orbit in a non-rotating (Schwarzschild) BH background was presented in~\cite{Vicente:2025gsg}. Recently, these results were extended to a spinning secondary in the Schwarzschild background~\cite{Karydas:2025bkj}. Here, we extend the results to a Kerr background, assuming that the secondary body is non-spinning. As we will see, this introduces non-trivial behavior in the friction force due to  relativistic frame dragging near the central BH. 

\subsection{Orbits in Kerr}

Let us first establish notation by introducing the background Kerr metric~\cite{Kerr:1963ud}. This metric describes the spacetime around a stationary, spinning BH with mass $M$ and angular momentum $J=aM^2$, where $a\in(-1,1)$ is the dimensionless Kerr parameter. 
In Boyer-Lindquist coordinates, the line element in this background reads
\begin{equation}
\begin{aligned}
\label{eq:Kerr_metric}
\dd s^2=&-\left(1-\frac{2r}{\Sigma}\right)\dd t^2+\frac{\Sigma}{\Delta}\dd r^2-\frac{4ar\sin^2\theta}{\Sigma}\dd t\dd\varphi\\
&+\Sigma \dd \theta^2+\left(r^2+a^2+\frac{2ra^2\sin^2\theta}{\Sigma}\right)\sin^2\theta\dd\varphi^2,
\end{aligned}
\end{equation}
with $\Sigma=r^2+a^2\cos^2\theta$, and $\Delta=r^2-2r+a^2$. We now assume that a secondary body is moving along a geodesic $\bm{\xi}$ in the spacetime metric of~\eqref{eq:Kerr_metric}, with 4-velocity $\bm{u}=\dd\bm{\xi}/\dd\tau$, where $\tau$ is the body's proper time. 
The 4-velocity in the Kerr background is uniquely determined by the energy $\varepsilon$, Carter constant $C$, and angular momentum about the BH's axis of rotation, $\ell_z$. Given a specific geodesic trajectory, these quantities read 
\begin{equation}
\varepsilon=-\bm{u}\cdot\bm{\partial}_t,~~~C=K(\bm{u},\bm{u}), ~~~\ell_z=\bm{u}\cdot\bm{\partial}_\varphi,
\end{equation}
where $\bm{\partial}_t,\bm{\partial}_\varphi$ are the timelike and azimuthal Killing vectors of the Kerr spacetime, and $K(\bm{u},\bm{u})$ is the second order Killing-Yano tensor in Kerr~\cite{Wald:1984rg}. 
At adiabatic order, the secondary's dynamics may be described as a sequence of geodesics whose flow is driven by gravitational radiation reaction. As the secondary moves, global energy and angular momentum fluxes are released into the primary and out to spatial infinity, which in turn causes the constants of motion associated with the trajectory of the secondary to change. 

In an EMRI, these fluxes can be computed for a given orbital trajectory using BH perturbation theory. 
At linear order in vacuum, the gravitational response of the system is described by the Teukolsky equation 
with a source term encoding the properties of 
the orbiting body.
In the presence of an environment, however, the Teukolsky equation is modified due to the different background geometry alongside additional environmental perturbations which may couple to the gravitational sector. Recent progress has been made towards fully relativistic Teukolsky evolutions in a non-vacuum background using two-parameter expansions which treat the environment as a perturbatively small correction to the background Kerr solution~\cite{Dyson:2025dlj,Dyson:2026ddd,Li:2025ffh}. Such approaches are appropriate for diffuse media and generally result in small global flux corrections, but it is unclear if they adequately capture the local dynamical interactions of the secondary with its environment. Recent efforts have shown that frictional effects may be captured self-consistently within gravitational self-force methods~\cite{Datta:2026eqj}, although more work is needed to properly achieve non-vacuum, self-force driven EMRI evolutions. Effective field theory methods, in which one computes local dynamical interactions between an orbiting body and the environment, offer a promising alternative route to a self-consistent treatment of non-vacuum EMRIs~\cite{Wilson-Gerow:2026plu,Mitra:2026gar,Modrekiladze:2024htc,Modrekiladze:2026twz}, but more progress is needed to incorporate the back-reaction of the perturber onto the fluid and perform full-blown EMRI evolutions.

For our present purposes, we assume that the DM halo is diffuse enough that metric back-reaction is negligible, and that the GW emission obeys the standard Teukolsky equation in vacuum with source at leading order. For a Milky-Way like DM halo with mass $\sim10^{12}M_\odot$ spread out over a $\sim20\text{kpc}$ spherical radius, this ``diffuse'' assumption holds. The effect of the environment on the secondary is encoded in a local 4-acceleration $\bm{\mathcal{A}}$, defined along a geodesic trajectory as $\bm{\mathcal{A}}=\nabla_{\bm{u}}\bm{u}$. In addition to gravitational radiation reaction, this acceleration affects the constants of motion according to~\cite{Hughes:1999bq}
\begin{equation}
\label{eq:flux}
\dot{\varepsilon}_{\rm{e}}=-\frac{1}{u^t}\bm{\mathcal{A}}\cdot\bm{\partial}_t,~~~\dot{\ell}_{z,\rm{e}}=\frac{1}{u^t}\bm{\mathcal{A}}\cdot\bm{\partial}_\varphi,~~~\dot{C}_{\rm{e}}=\frac{2}{u^t}K(\bm{\mathcal{A}},\bm{u}).
\end{equation}
Here, the subscript ``$\rm{e}$'' stands for ``environment,'' and the overdot denotes a derivative with respect to coordinate time $t$. The expressions in Eq.~\eqref{eq:flux} are valid for a generic instantaneous 4-acceleration of the secondary body. In this work, we are interested in computing the accelerations due to scattering and accretion of an ensemble of collisionless particles. We turn to this now.  

\subsection{Phase space in the Kerr background} 
We assume that the background is a non-dissipative, collisionless halo of DM, made up of particles with rest mass $\mu$ (see~\cite{Ostriker:1998fa,Kim:2007zb,Barausse:2007ph} for treatments of dynamical friction in dissipative and collisional media such as dense gas environments in less relativistic contexts).

As the secondary body moves through the DM, gravitational capture (accretion) and scattering events cause a momentum transfer between the medium and the secondary body. In intermediate-mass-ratio inspirals (IMRIs), DM scattering may induce a non-negligible change in the DM halo's phase space distribution throughout the inspiral due to feedback effects~\cite{Kavanagh:2020cfn,Coogan:2021uqv,Nichols:2023ufs,Kavanagh:2024lgq,Wade:2025rkk,Wade:2026xvm}. Here, our primary concern is with EMRI systems where such effects have been shown to be negligible, although this point has been recently debated when considering the impact of three-body slingshot effects and more realistic astrophysical models ~\cite{Sharpe:2026nqq,Sharpe:2026kfk,Karydas:2026ctq}. Nevertheless, the methods we present could be used in conjunction with an evolving phase space with little conceptual difficulty.

First, we consider the differential number of particles in a phase space volume $\dd V\dd P$, given by $\dd N=\left(-\bm{p}\cdot\bm{n}\right)F(\bm{x},\bm{p})\dd V\dd P$~\cite{Lindquist:1966igj}, where $\bm{p}=\mu\bm{w}$ is the 4-momentum of the particles in the distribution, $\bm{w}$ their 4-velocity, and $\mu$ the mass of an individual particle. Throughout, we assume that the collisionless medium consists entirely of a single particle type with mass $\mu$. The vector $\bm{n}$ is a forward-directed timelike unit vector which is orthogonal to the spatial directions $\dd V$, and can be freely chosen depending on our frame of interest. 
As defined here, $F(\bm{x},\bm{p})$ is the invariant phase space distribution function. The momenta in the phase space are further constrained to be within the mass shell, i.e., the only allowed momenta are the set of vectors $\bm{p}$ with fixed length $\bm{p}\cdot\bm{p}=\mu^2(\bm{w}\cdot\bm{w})=-\mu^2$. Assuming the DM particle distribution is in equilibrium, the distribution function may be written solely as a function of the constants of motion 
$(\mathcal{E},\mathcal{C},\mathcal{L}_z)$~\cite{10.1093/mnras/76.2.70}, which allows us to define  $F(\bm{x},\bm{p})=\mu^{-3}f(\mathcal{E},\mathcal{C},\mathcal{L}_z)$.

Now, let us pick a set of frame vectors centered on the secondary body. We define a tetrad basis $(\bm{\partial}_0,\bm{\partial}_i)=(\bm{u},\bm{\partial}_i)$, and choose the spatial vectors in such a way that the frame is locally orthonormal. With this choice, the momentum space volume $\dd P$ can be written as $\dd P=(-1/p_0)dp^1dp^2dp^3$, 
where the differential momenta are chosen to lie along the spatial directions. We further define the DM particle 3-velocities through the relation $\bm{p}=\mu\gamma_v(1,v^i)$, where $\gamma_v=(1-|v|^2)^{-1/2}$, and $|v|^2=v^iv_i$. Swapping 4-momenta with 3-velocities, the momentum volume element becomes $\dd P=\mu^{2}\gamma_v^4dv^1dv^2dv^3=\mu^2\gamma_v^4\dd^3v$. Choosing $\bm{n}=\bm{u}$ such that $-\bm{p}\cdot\bm{n}=\mu\gamma_v$ and putting all the pieces together, the differential number of particles in a spatial volume $\dd V$ with given differential velocities $dv^1dv^2dv^3$ in the secondary body's local plane of simultaneity is given by~\cite{Vicente:2025gsg,Lindquist:1966igj}

\begin{equation}
\label{eq:Distribution_Function_2nd_frame}
\frac{\dd N}{\dd V\dd^3v}=\gamma_v^5f(\mathcal{E},\mathcal{C},\mathcal{L}_z).
\end{equation}

Note that Eq.~\eqref{eq:Distribution_Function_2nd_frame} describes the distribution function as seen by the secondary body, which is encoded primarily in the choice of $\bm{n}$. For example, in Section~\ref{sec:results} we will compare our relativistic results with classical and phenomenological models of DF. This requires the choice $\bm{n}=\{(r^2+a^2)^2-\Delta a^2\sin^2\theta\}^{1/2}\left(\Delta \Sigma\right)^{-1/2}\left(\bm{\partial}_t+\omega_K\bm{\partial}_\varphi\right)$, which is the 4-velocity of the ``zero angular momentum observers'' (ZAMOs), where the Kerr frequency is $\omega_K=-g_{t\varphi}/g_{\varphi\varphi}$. 
In the $a\to 0$ limit, this frame reduces to the frame of a static observer in the Schwarzschild background, $\bm{n}=(-g_{tt})^{-1/2}\bm{\partial}_t$. We will also make occasional use of the ``freely-falling frame'', or, the frame of the DM particles, where $\bm{n}=\bm{w}$, to compare with the results of Ref.~\cite{Ferrer:2017xwm} which computed the DM spike rest density in that frame. 

Now, particles on bound orbits in the Kerr geometry must satisfy several conditions which limit the allowed phase space region. Consider first a DM particle with 4-velocity $\bm{w}$. We may write its orbital constants of motion as
\begin{equation}
\begin{aligned}
\label{eq:DM_orbital_constants}
&~~~~~~~\mathcal{E}=-w_t,~~~~~\mathcal{L}_z=w_\varphi,\\
\mathcal{C}&=w_\theta^2+\frac{\mathcal{L}_z^2}{\sin^2\theta}+a^2\cos^2\theta(1-\mathcal{E}^2).
\end{aligned}
\end{equation}
Here, we have adopted a definition of the Carter constant in line with~\cite{Sadeghian:2013laa,Ferrer:2017xwm}, which reduces to the total angular momentum per unit mass (squared) $L^2$ in the $a\rightarrow0$ limit. In the background metric~\eqref{eq:Kerr_metric}, the radial and angular components $w^r,w^\theta$ satisfy the equations
\begin{equation}
w_\theta=\pm\sqrt{U(\theta)},~~~~~~~w_r=\pm\frac{r^2}{\Delta}\sqrt{V(r)},
\end{equation}
where the radial and angular potentials $V(r)$, $U(\theta)$ are
\begin{equation}
\label{eq:potentials}
\begin{aligned}
V(r)&=\left(1+\frac{a^2}{r^2}+\frac{2a^2}{r^3}\right)\mathcal{E}^2-\frac{\Delta}{r^2}\left(1+\frac{\mathcal{C}}{r^2}\right)\\
&~~~~~+\frac{a^2\mathcal{L}_z^2}{r^4}-\frac{4a\mathcal{E}\mathcal{L}_z}{r^3},\\
U(\theta)&=\mathcal{C}-\frac{\mathcal{L}_z^2}{\sin^2\theta}-a^2\cos^2\theta(1-\mathcal{E}^2).
\end{aligned}
\end{equation}
Assuming our system is in equilibrium, the only particles that will contribute to friction and accretion are particles on bound orbits around the primary BH, which are non-plunging. For a given $(\mathcal{E},\mathcal{C},\mathcal{L}_z)$ to correspond with a bound orbit, the following conditions must be satisfied (see~\cite{Ferrer:2017xwm} for more details). First, $\mathcal{E}\leq1$, $V(r)\geq0$, and $U(\theta)\geq0$. Next, to exclude the plunging orbits at a given $r$, there must be a potential barrier preventing particles from falling into the horizon. This means that a minimum in the potential at a radius $r_{\rm{unst}}$ must exist outside the horizon $r_{\rm{H}}$ such that $V'(r_{\rm{unst}})=0$, with $r_{\rm{H}}\leq r_{\rm{unst}}\leq r$. The point $r_{\rm{unst}}$ must exist in the forbidden region $V(r_{\rm{unst}})<0$ to ensure that there is indeed a barrier between the point $r$ and the horizon, preventing particles from falling into the central BH. These conditions define the loss cone of the distribution, and ensure that only non-plunging bound particles are included in the phase space distribution function $f(\mathcal{E},\mathcal{C},\mathcal{L}_z)$. Enforcing these conditions on the phase space yields a generically anisotropic distribution function, which we discuss in detail below.

\subsection{Friction and accretion for circular equatorial orbits}
\label{sec:friction_accretion}

We now have all the pieces we need to compute the effects of friction and accretion on an orbiting body. Specializing to circular equatorial orbits, such that the secondary body's 4-velocity components are $u^r=u^\theta=0$, and $\theta=\pi/2$. 
We choose our frame vectors such that 

\begin{equation}
\label{eq:frame_vectors}
\begin{aligned}
\bm{\pd}_0&=\bm{u}=u^t\bm{\pd}_t+u^\varphi\bm{\pd}_\varphi,\\
\bm{\pd}_1&=\frac{\sqrt{\Delta}}{r}\bm{\pd}_r,\\
\bm{\pd}_2&=\frac{1}{r}\bm{\pd}_\theta,\\
\bm{\pd}_3&=\frac{1}{\sqrt{g_{\varphi\varphi}+\ell_z^2}}\left(\bm{\pd}_\varphi+u_\varphi\bm{u}\right).
\end{aligned}
\end{equation}
We can easily relate the DM particles' orbital constants of motion to their 3-velocities in the frame defined by Eq.~\eqref{eq:frame_vectors}, recalling Eqs.~\eqref{eq:DM_orbital_constants} and noting that $\bm{w}=\gamma_v(\bm{\partial}_0+v^i\bm{\partial}_i)=w^\mu\bm{\partial}_\mu$. We obtain 
\begin{equation}
\label{eq:ECLz_secondary_frame}
\begin{aligned}
\mathcal{E}&=\gamma_v\left(\varepsilon+v^3\Omega\sqrt{g_{\varphi\varphi}+\ell_z^2}\right),\\
\mathcal{L}_z&=\gamma_v\left(\ell_z+v^3\sqrt{g_{\varphi\varphi}+\ell_z^2}\right),\\
\mathcal{C}&=\mathcal{L}_z^2+\left(r\gamma_v v^2\right)^2,\\
\end{aligned}
\end{equation}
where $\varepsilon=\frac{a+\sqrt{r}(r-2)}{\sqrt{r^2(r-3)+2ar^{3/2}}}$ and $|\ell_z|=\frac{r^2-2a\sqrt{r}+a^2}{\sqrt{r^2(r-3)+2ar^{3/2}}}$ 
are the secondary BH's specific energy and angular momentum (normalized by the primary mass), respectively, and $\Omega=\frac{\dd \varphi}{\dd t}=u^\varphi/u^t=1/(r^{3/2}+a)$ is the secondary's azimuthal orbital frequency. Taking the expressions in Eq.~\eqref{eq:ECLz_secondary_frame} alongside the conditions for bound, non-plunging orbits, one can see that the distribution function of Eq.~\eqref{eq:Distribution_Function_2nd_frame} will generically depend on the specific velocity components $(v^2,v^3)$, and hence is anisotropic. This holds true even for constant distribution functions $f=f_0$, as the anisotropy is tied to the secondary body's frame and the loss cone in the Kerr geometry, rather than properties of the distribution.

Next, in the frame specified by Eqs.~\eqref{eq:frame_vectors}, we wish to compute the acceleration on the secondary body due to particle scattering and accretion. The acceleration $\bm{\mathcal{A}}=\nabla_{\bm{u}}\bm{u}$ satisfies the equation 
\begin{equation}
\label{eq:acceleration_general}
    \bm{\mathcal{A}}=\frac{1}{m}\left[\nabla_{\bm{u}}(m\bm{u})-\bm{u}\frac{\dd m}{\dd \tau}\right],
\end{equation}
where $m$ is the mass of the secondary.
The first term encodes the transfer of momentum from the particles to the secondary, with the second term describing the secondary body's mass accretion due to particle capture. Assuming the effect of particle interactions taking place in the ``body-zone'' of the secondary, and further noting that the secondary's mass change is a higher order effect compared to the scattering and accretion momentum transfer,
the acceleration Eq.~\eqref{eq:acceleration_general} may be evaluated in Fermi normal coordinates as~\cite{Vicente:2025gsg}
\begin{equation}
\label{eq:acceleration_DF_ACC}
\bm{\mathcal{A}}=\frac{1}{m}\int\frac{\dd N}{\dd V \dd^3v}|v|\left[\sigma_{\rm{scat}}+\sigma_{\rm{acc}}\right]\bm{p}_\perp\dd^3v.
\end{equation}
Here, $\bm{p}_\perp=\mu\gamma_v(0,v^i)$ is the perpendicular component of the incoming particle momenta, $(\dd N/\dd V \dd^3v)$ is the number density of DM particles per unit spatial and phase space volume, given by Eq.~\eqref{eq:Distribution_Function_2nd_frame}, and $\sigma_{\rm{scat}},\sigma_{\rm{acc}}$ are the scattering and accretion cross sections, respectively. When the secondary body is a Schwarzschild BH, the cross section for particle capture can be solved exactly in terms of the critical impact parameter  as $\sigma_{\rm{acc}}=\pi b_{\rm{cr}}^2$, with $b_{\rm{cr}}$ given by 
\begin{equation}
\label{eq:critical_impact_parameter}
b_{\rm{cr}}^2/m^2=\frac{-1+20|v|^2+8|v|^4+\left(1+8|v|^2\right)^{3/2}}{2|v|^4}.
\end{equation}
The scattering cross section can be defined as an integral over impact parameters greater than the critical capture parameter $b_{\rm{cr}}$, i.e., $\sigma_{\rm{scat}}=4\pi m^2\int_{b>b_{\rm{cr}}}\dd b ~b\cos^2\left[\phi_\infty(b,|v|)\right]$. The quantity $\phi_\infty$ is the asymptotic true anomaly, related to the deflection angle as $\pi-2\phi_\infty$, which can be expressed in terms of elliptic integrals of the first kind~\cite{Traykova:2021dua,Chandrasekhar:1985kt}. Relativistic corrections to the scattering in the strong field regime are encoded in the parameter $b_{\rm{gr}}$, defined as $b_{\rm{gr}}/m=\left(1+|v|^2\right)/|v|^2$. The scattering integral in general must be evaluated numerically between the critical impact parameter $b_{\rm{cr}}$ and some chosen maximum impact parameter $b_{\rm{max}}$, which we must choose appropriately to avoid divergences. Here, we assume that the body-zone of the secondary mass extends to its Hill radius $r_{\rm{Hill}}=r(q/3)^{1/3}$, where $q=m/M$ is the mass ratio, 
so that the particles within $b_{\rm{max}}=r_{\rm{Hill}}$ are the only ones that contribute significantly to the scattering. Other physically meaningful choices can be made depending on the extent to which long-range particle interactions contribute; the Hill radius can be a bad choice for models of friction in diffuse, low-collisional accretion disks, for example, since long range interactions with propagating density perturbations in the disk well outside the Hill sphere contribute significantly to resonant torques on the secondary~\cite{Goldreich:1979zz,Goldreich:1980wa,Ward:1997di}. As defined in~\cite{Vicente:2025gsg}, we can write the scattering cross section more explicitly as $\sigma_{\rm{scat}}=4\pi m^2b_{\rm{gr}}^2\Lambda$, where $\Lambda$ is the Coulomb logarithm, defined as
\begin{equation}
\Lambda(v)=\ln\sqrt{\frac{\bmax^2+\bgr^2}{\bmin^2+\bgr^2}}+\bgr^{-2}\int_{\bcr}^{\bmin}\dd b~b\cos^2\left[\phi_\infty(b,|v|)\right].
\end{equation}
The minimum impact parameter $\bmin$ can be freely chosen provided that $\bcr\ll\bmin\leq\bmax$. Generally, the integral must be evaluated numerically, but it can be approximated with a fitting function given in the Supplemental Material of Ref.~\cite{Vicente:2025gsg}. 

Using Eqs.~\eqref{eq:Distribution_Function_2nd_frame} and~\eqref{eq:ECLz_secondary_frame} alongside expressions for the scattering and accretion cross sections $\sigma_{\rm{scat}}$ and $\sigma_{\rm{acc}}$, we can numerically integrate the acceleration integral Eq.~\eqref{eq:acceleration_general} for a given distribution function $f(\mathcal{E},\mathcal{C},\mathcal{L}_z)$. Since we are ultimately interested in EMRI evolutions within DM halos, we can then use Eqs.~\eqref{eq:flux} to relate the acceleration in the secondary body's frame to the change in the secondary body's orbital constants of motion. For circular equatorial orbits, the expressions can be written in terms of the acceleration components as 
\begin{equation}
\begin{aligned}
\dot{\varepsilon}_{\rm{e}}&=\mathcal{A}^3\Omega\frac{\sqrt{g_{\varphi\varphi}+\ell_z^2}}{g^{t\varphi}\ell_z-g^{tt}\varepsilon}=\mathcal{A}^3\Omega\sqrt{\Delta},\\
\dot{\ell}_{z,\rm{e}}&=\mathcal{A}^3\frac{\sqrt{g_{\varphi\varphi}+\ell_z^2}}{g^{t\varphi}\ell_z-g^{tt}\varepsilon}=\mathcal{A}^3\sqrt{\Delta}.
\\\dot{C}_{\rm{e}}&=\frac{2\mathcal{A}^3\left(\ell_z-a\varepsilon\right)}{\sqrt{g_{\varphi\varphi}+\ell_z^2}}\left((r^2+a^2)\varepsilon-a\ell_z\right)\\
&=\mathcal{A}^3\frac{2\left(r^2-ar^{3/2}\right)\Omega\sqrt{\Delta}}{\sqrt{1-3/r+2a/r^{3/2}}}.
\end{aligned}
\end{equation}

\begin{figure*}[t]
\centering 
\includegraphics[width=\linewidth]{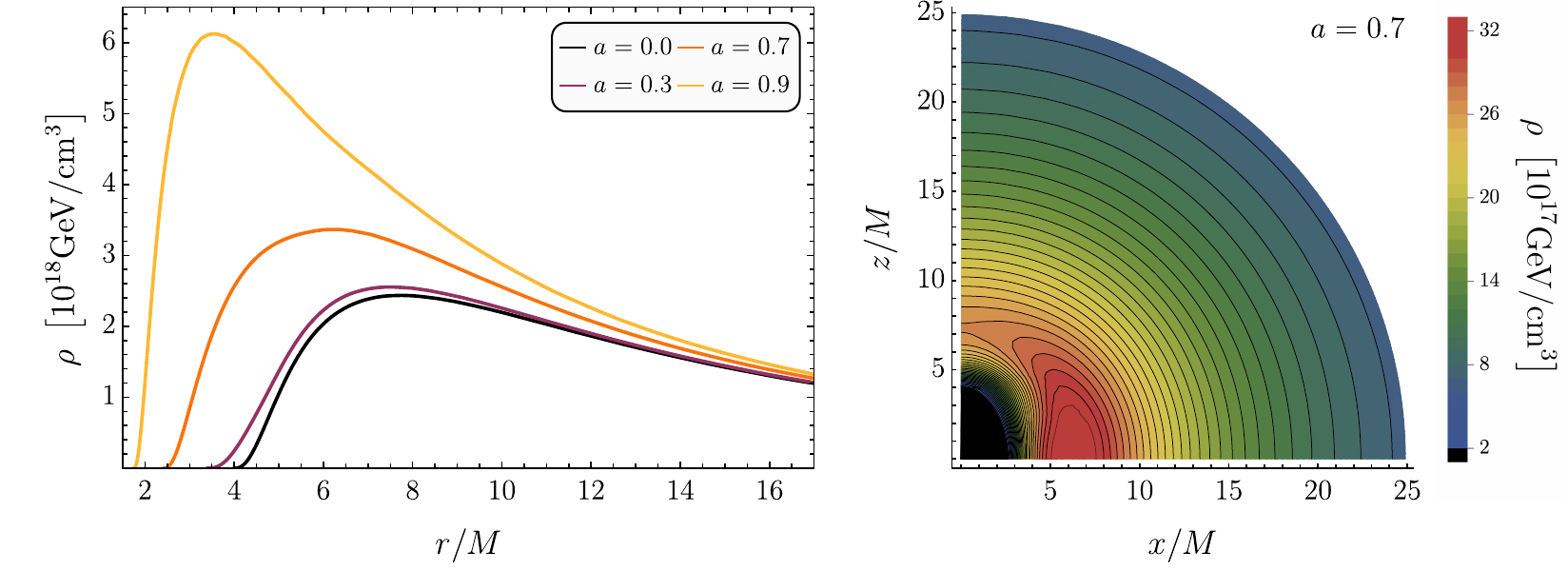}
\caption{Left: Adiabatically grown DM 
density distributions on the $\theta=\pi/2$ plane as seen in the fluid's freely falling ($\boldsymbol{n}=\boldsymbol{w}$) frame, for various choices of the central BH spin. Right: Angular slice of the final DM distribution for a particular choice of BH spin, plotted in terms of the flat space spherical coordinates $x,z$. In both panels, we assume an initial Hernquist distribution with mass $M_{\text{H}}=10^{12}M_\odot$ and scale radius $r_s=20\text{kpc}$, and grow the distribution around a central BH with mass $M=4\times10^6M_\odot$. The inner edge of the density corresponds to the prograde marginally bound orbit.
}
\label{fig:DM_density}
\end{figure*}

\section{Collisionless Dark Matter Halos in a Kerr Background}\label{sec:adiabatic_growth}
Equations~\eqref{eq:frame_vectors},~\eqref{eq:ECLz_secondary_frame}, and~\eqref{eq:acceleration_general} in Section~\ref{sec:friction_accretion} are generically valid for a small body orbiting a central Kerr BH surrounded by a distribution of DM. In principle, we can plug in any distribution function we wish into Eq.~\eqref{eq:acceleration_general} to obtain the secondary's acceleration. To obtain physically meaningful results which are valid in a stationary background in equilibrium, however, we should also impose the loss cone conditions $V\geq0$, $U\geq0$, $r_\text{H}\leq r_{\rm{unst}}\leq r$, etc. when integrating over the velocities $\dd^3v=d v^1d v^2d v^3$. 
One approach which ensures the loss cone conditions are satisfied, represents a physically meaningful distribution of DM, and is valid in the relativistic regime of Kerr, is to consider the DM halo's adiabatic response to the growth of a central BH.

The adiabatic growth of a DM halo in the vicinity of a central supermassive BH has been studied extensively in the literature. Originally, Gondolo and Silk studied the growth of such DM ``spikes'' by considering the adiabatic evolution of a DM halo as a central point mass is grown in a Newtonian framework~\cite{Gondolo:1999ef}. Later efforts placed the computation on a fully relativistic footing in both Schwarzschild and Kerr backgrounds~\cite{Sadeghian:2013laa,Ferrer:2017xwm}, revealing interesting differences from the Newtonian calculation in the strong field regime. Here, we review essential pieces of the relativistic computation in Kerr. For more details, we refer the reader to~\cite{Ferrer:2017xwm}.

\subsection{Adiabatic growth in the Kerr background}

We start by considering the DM halo's description in phase space before the central BH grows. Assuming that the initial density distribution $\rho(r)$ is spherically symmetric and that the distribution is ergodic, the phase space distribution function is solely a function of the particle energy, $f=f(E)$. With a realistically diffuse halo of DM, its gravitational potential $\Phi(r)$ is well approximated by Newtonian gravity, and we may use this fact to obtain $f(E)$ for a given initial $\rho(r)$ through standard Eddington inversion methods~\cite{Lacroix:2018qqh}.

Now, we consider the adiabatic growth of a central Kerr BH within the DM distribution. This growth preserves the functional form of $f$ (see, e.g., Appendix A of~\cite{Ferrer:2017xwm}), so that the initial non-relativistic distribution function $f(E,L,L_z)$ can be related to the final relativistic distribution after the BH grows, $f(\mathcal{E},\mathcal{C},\mathcal{L}_z)$. Here, $(E,L,L_z)$ are the DM particle energy, total angular momentum, and angular momentum about the $z$-axis, respectively, with $(\mathcal{E},\mathcal{C},\mathcal{L}_z)$ the particle energy per unit mass, Carter constant per unit mass squared, and angular momentum about the $z$-axis per unit mass in the Kerr background. The adiabatic growth shifts the DM particle energies and angular momenta as the potential changes in such a way that the action variables $I_r$, $I_\theta$, and $I_\phi$ are invariant~\cite{1980ApJ...242.1232Y,Sadeghian:2013bga}. In the initial non-relativistic distribution, the action integrals read 
\begin{equation}
\begin{aligned}
\label{eq:Halo_Invariants_pre_growth}
I_r'(E,L)&=\oint\dd r\sqrt{2E-2\Phi-L^2/r^2},\\
I_\theta'(L,L_z)&=\oint\dd \theta\sqrt{L^2-\frac{L_z^2}{\sin^2\theta}}=2\pi\left(L-|L_z|\right),\\
I_\phi'(L_z)&=\oint\dd\phi L_z=2\pi L_z,
\end{aligned}
\end{equation}
where $\Phi$ is the Newtonian gravitational potential of the DM halo.
After the BH growth, for particles on bound orbits in the Kerr background, the action integrals read 
\begin{equation}
\begin{aligned}
\label{eq:Halo_Invariants_post_growth}
I_r(\mathcal{E},\mathcal{C},\mathcal{L}_z)&=\oint u_r\dd r=\oint\dd r\frac{\sqrt{V(r)}}{1-2/r+a^2/r^2},\\
I_\theta(\mathcal{E},\mathcal{C},\mathcal{L}_z)&=\oint u_\theta\dd \theta=\oint\dd \theta\sqrt{U(\theta)},\\
I_\phi(\mathcal{L}_z)&=\oint u_\phi\dd \phi=2\pi\mathcal{L}_z,
\end{aligned}
\end{equation}
where the potentials $V(r)$ and $U(\theta)$ are defined in Eq.~\eqref{eq:potentials}. To obtain a given point in the new phase space, we can relate the action integrals before and after growth to obtain $(E,L,L_z)$ as functions of the orbital parameters $\left(\mathcal{E},\mathcal{C},\mathcal{L}_z\right)$. This procedure yields the post-growth distribution function $f(\mathcal{E},\mathcal{C},\mathcal{L}_z)$, which can be further written as a function of the DM particle 3-velocities in the secondary's frame via  Eq.~\eqref{eq:ECLz_secondary_frame}. Example density distributions in the fluid rest frame ($\boldsymbol{n}=\boldsymbol{w}$) are shown in Fig.~\ref{fig:DM_density}, for a DM halo with total mass $M_{\text{H}}=10^{12}M_\odot$, scale radius $r_s=20\text{kpc}$, and various central BH spin values.

We note that the trajectory of 
the secondary introduces a 3-velocity 
dependence on the DM particles' orbital parameters, through the choice of the body-centered orthonormal frame vectors. This induces an anisotropy in the distribution function which is unavoidable, as the loss cone conditions in the Kerr background become dependent on specific velocity components of the DM. As pointed out in Ref.~\cite{Vicente:2025gsg}, this anisotropy is unaccounted for in prescriptions of DF which adopt modified Chandrasekhar-like formulae for the secondary body's energy loss. Indeed, ignoring such anisotropy may lead to an over-estimation of the friction force due to an over-counting of scattering contributions to the integral Eq.~\eqref{eq:acceleration_DF_ACC}, as we demonstrate explicitly in Sec.~\ref{sec:results}.

\section{Results}
\label{sec:results}
With the phase space considerations on the Kerr background, prescription for swapping to the secondary's locally Minkowski frame, and mapping between the DM particles' local velocities $v^i$ 
to the orbital constants of motion in Eq.~\eqref{eq:ECLz_secondary_frame} given in the previous section, we have all the pieces necessary to compute the acceleration integral of Eq.~\eqref{eq:acceleration_DF_ACC}. In this section, we will show the relativistically computed friction and accretion on a secondary body within the adiabatically grown DM distribution discussed in Sec.~\ref{sec:adiabatic_growth}. Throughout this section 
we take the background DM to be grown from an initial spherically symmetric Hernquist distribution whose distribution function $f(E)$ is ergodic, and can be computed in closed form via Eddington inversion (see, e.g.,~\cite{Sadeghian:2013laa,Ferrer:2017xwm,Lacroix:2018qqh}). We will compare these results to classical DF prescriptions, and discuss where the computations differ. 
\subsection{Friction and accretion in the Kerr background}
\label{sec:DF_acc_in_Kerr}
\begin{figure}
    \centering
    \includegraphics[width=\linewidth]{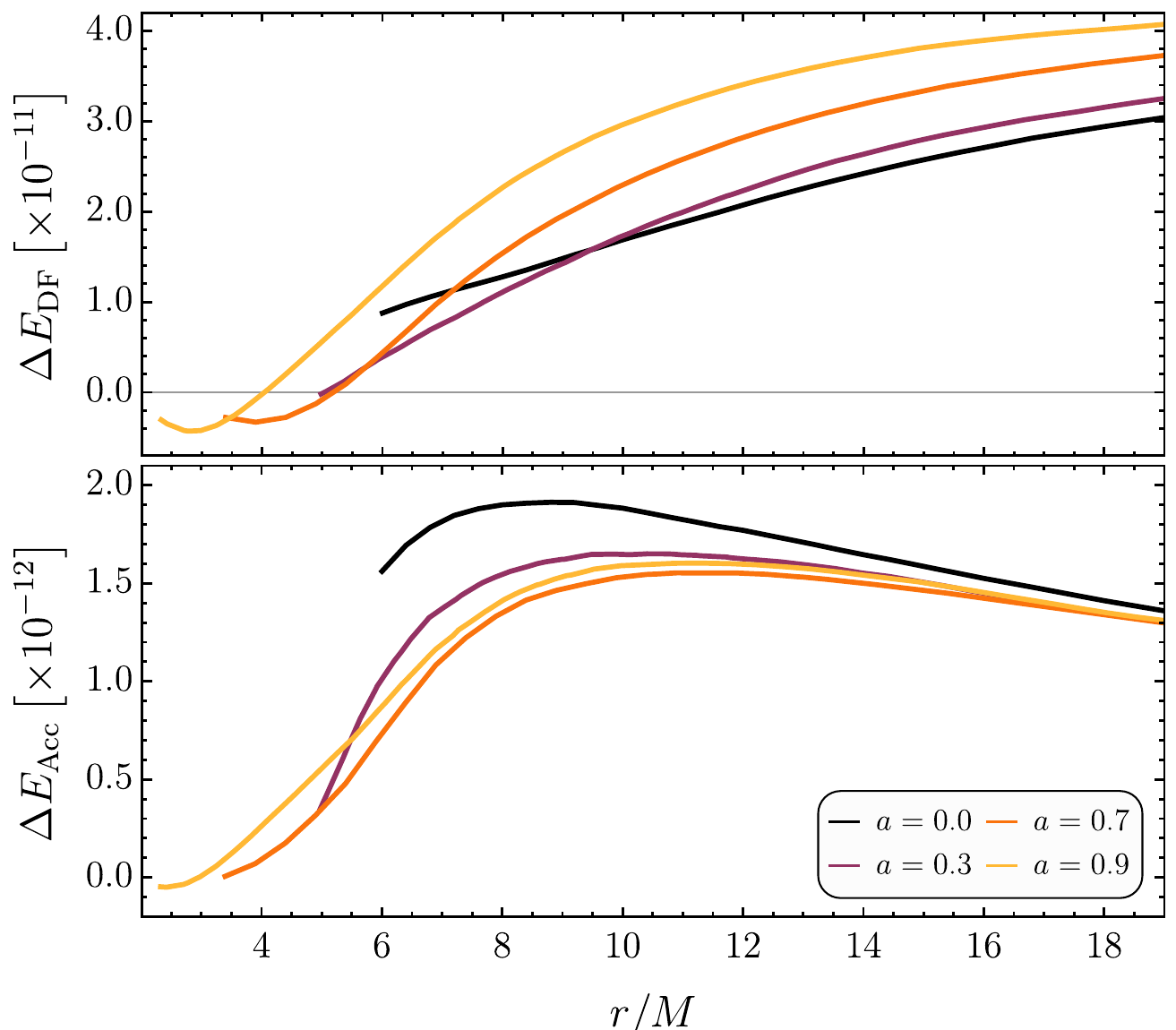}
    \caption{Top: Energy loss during one orbit due to scattering events for various choices of central BH spin. Bottom: Same as the top panel, for particle momentum accretion. We fix the total mass and scale radius of the DM halo to be $M_{\text{H}}=10^{12}M_\odot$ and $r_s=20\text{kpc}$, respectively, and take the primary and secondary masses to be $M=10^6M_\odot$ and $m=10M_\odot$. The results are shown for circular equatorial orbits, and lines are cut off when the secondary reaches its prograde ISCO.}
    \label{fig:DF_acc_Kerr}
\end{figure}

In Fig.~\ref{fig:DF_acc_Kerr}, we show the energy losses due to DF (top) and accretion (bottom) for a small mass moving along circular orbits in Kerr backgrounds with selected values of the spin. In this configuration, accelerations $\mathcal{A}^1$ and $\mathcal{A}^2$ computed through Eq.~\eqref{eq:acceleration_DF_ACC} vanish. Physically speaking, there is no net radial or polar flow of the DM particles for a given point in the equatorial plane for a stationary axisymmetric DM distribution, and hence no net accelerations in these directions since the secondary moves only along the $\boldsymbol{\partial}_\varphi$ direction. We compute the energy loss $\Delta E$ over one orbital period with time $T=2\pi/\Omega$ as $\Delta E=\dot{\varepsilon}(2\pi/\Omega)$, with $\dot{\varepsilon}$ computed via Eq.~\eqref{eq:flux}. The background DM distribution has the same total mass and scale radius as in Fig.~\ref{fig:DM_density}. 
The primary and secondary masses are chosen to be  $M=10^6M_\odot$ and $m=10M_\odot$, respectively. The secondary's orbit is circular, equatorial ($\theta=\pi/2$), and prograde with respect to the primary's spin.   

At larger distances, frictional effects become more prominent as the central BH spin is increased. The far region approaches the Newtonian regime, where one expects the DF force to be $\propto\rho(r)$. At a given radius $r/M$, higher values of the central BH spin correspond to higher values of the density distribution on the equatorial plane, and hence the large distance, weak field regime approaches the Newtonian expectation. 
However, things are different in the strong field regime. 
For higher spin values, as the secondary body approaches the ISCO, we see a decrease and eventual sign flip in the energy lost to DF.
This effect cannot be captured by Newtonian or relativistically-corrected prescriptions which assume that the friction is proportional to the local density. Indeed, for highly spinning primaries, the density distributions in Fig.~\ref{fig:DM_density} reach their highest values in the region where frictional effects weaken and change sign. Accretion onto the secondary is impacted by the primary's spin as well, albeit with a much milder dependence on $a$. Energy losses due to accretion generally decrease as the spin is increased, although the behavior is not monotonic.

The behaviors in Fig.~\ref{fig:DF_acc_Kerr} may be understood as follows. First, we note that particles in the background have prograde and retrograde marginally bound orbits at radii $r_{\text{mb}}^{\pm}=2\mp a+2\sqrt{1\mp a}$ for a given BH spin, with $r_{\text{mb}}^+<r_{\text{mb}}^-$. 
The location of $r_{\text{mb}}^+$ defines the inner edge of the density distributions considered here, as seen in Fig.~\ref{fig:DM_density}. In the region $r_{\text{mb}}^+<r<r_{\text{mb}}^-$, the only stable orbits of the DM particles are prograde trajectories for which $\mathcal{L}_z>0$. For low spins, the secondary body reaches its prograde ISCO radius $r_{\text{ISCO}}^+$ before entering the $r_{\text{mb}}^+<r<r_{\text{mb}}^-$ region. As the spin grows, $r_{\text{ISCO}}^+$ decreases while $r_{\text{mb}}^-$ increases: for $a\gtrsim0.3726$, we eventually have that $r_{\text{ISCO}}^+<r_{\text{mb}}^-$. 
When this condition holds, the secondary body can spend significant time orbiting within the region where the surrounding DM particles are all moving in the same direction. Deep in this region, the background particles can on average move faster than the secondary body, and hence induce an impulsive (rather than frictional) force on the secondary.
\begin{figure*}
\centering
\includegraphics[width=0.98\linewidth]{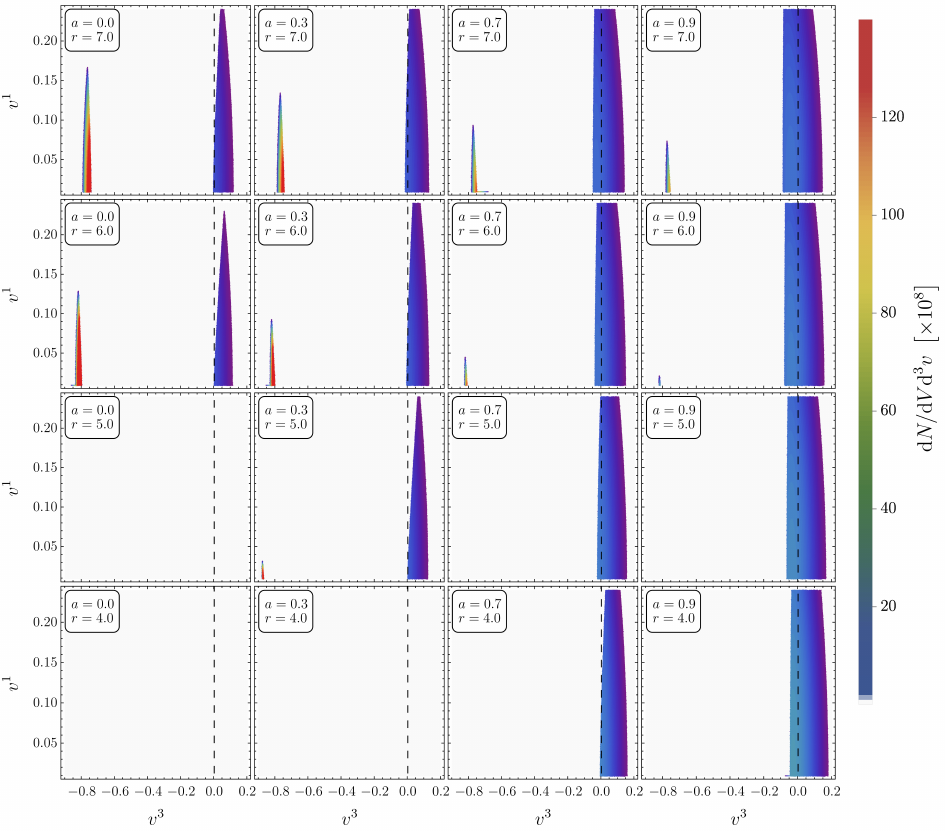}
\caption{Allowed phase space distribution of DM particles in the $v^2=0$ plane, as seen in the secondary's local frame defined by $\boldsymbol{n}=\boldsymbol{u}$. At large radii, particles moving with highly negative $v^3$ dominate the scattering. As the orbital radius decreases, $v^3>0$ contributions become more important and can eventually dominate at small enough radii for highly spinning primaries. 
}
\label{fig:phase_space}
\end{figure*}
This is shown explicitly in Fig.~\ref{fig:phase_space}, where we plot the allowed phase space distribution $\dd N/\dd V\dd^3v$ in the secondary body's local frame $\boldsymbol{n}=\boldsymbol{u}$, as a function of the orbital radius 
and primary spin. 
On the equator in the $v^2=0$ plane, the allowed phase space of the DM particles separates into two regions roughly centered on the (Newtonian) speeds of the co-rotating and counter-rotating circular orbits, $v^\pm_{\text{DM}}\sim \pm r/\left(r^{3/2}\pm a\right)$. In terms of $v^3$ in the secondary's frame, the phase space regions roughly center around co- and counter-moving orbits given by the relative speeds of the DM particles and secondary, $v^3\sim v_{\text{DM}}^+-r/\left(r^{3/2}+ a\right)=0$ and $v^3\sim v_{\text{DM}}^--r/\left(r^{3/2}+ a\right)=-2r^{5/2}/(r^3-a^2)$. For the DF force, contributions to the acceleration integral are dominated by particles moving with large negative $v^3$ at large radii, for all values of the spin. As the secondary moves closer to its ISCO, the overall negative contributions decrease as the allowed phase space region shrinks there, and eventually the positive-$v^3$ scattering contributions may become dominant. 

The location where the sign flips is non-trivially dependent on the choice of $b_{\text{max}}$: here, we have taken $b_{\text{max}}=r_{\text{Hill}}=r(q/3)^{1/3}$, which gives an explicit dependence on the mass ratio in the scattering cross-section. Other choices (e.g., $b_{\text{max}}\sim2r$, motivated by simulations of DF in scalar clouds~\cite{Traykova:2021dua,Traykova:2023qyv,Wang:2026daj}) may include a much larger range of scatterers, and hence differences in the location of the sign change. As we will demonstrate, the location of the sign flip and the magnitude of the friction force are also sensitive to the choice of the initial distribution function. The choice of $b_{\text{max}}$ warrants its own investigation, as recent work using effective field theory techniques shows that self-consistent choices of $b_{\text{max}}$ may arise naturally from the choice of orbit and global shape of the environment~\cite{Mitra:2026gar}. 

The trend in the accretion may be explained similarly. The portion of large, negative-$v^3$ DM particles drops for a given orbital radius as the spin is increased. The accretion cross section favors small velocities, hence the contributions to the acceleration from the $v^3\sim0$ region begin to dominate. As the primary spin increases, this $v^3\sim0$ region centers around $v^3=0$; positive and negative contributions to Eq.~\eqref{eq:acceleration_DF_ACC} cancel out, decreasing the accretion.

The behavior of the DF and accretion forces depends intimately on the type of secondary body~\cite{Karydas:2025bkj,Karydas:2025gqp}.
For secondary neutron stars and collisionless DM, the effect of small impact parameter scattering from DM which passes through the neutron star may disrupt the behavior seen in Fig.~\ref{fig:DF_acc_Kerr}. 
A detailed comparison of DF and accretion for different types of compact objects 
interacting with rotating collisionless backgrounds is an interesting topic for future work.

\subsection{Impact of the initial distribution function}
Cosmological $N$-body simulations indicate that galactic DM halos generally exhibit velocity anisotropy and nonzero net angular momentum. In particular, the mean specific angular momentum projected along the halo's total angular-momentum axis, averaged over spherical shells, has been found to approximately follow a power-law radial profile, \(\langle L_z\rangle(r)\propto r^\alpha\), with \(\alpha=1.1\pm0.3\) \cite{Barnes:1987hu,Bullock:2000ry}. Here, \(L_z\) denotes the specific angular momentum component along the halo spin axis. This empirical scaling characterizes the radial distribution of net angular momentum but does not uniquely determine the halo's phase-space distribution function.
In the context of galactic dynamics, a damping of the DF force acting on galactic bars has been seen to occur solely due to the uneven splitting of prograde and retrograde motion of DM particles in such $L_z$-dependent halos~\cite{Chiba:2023rqm}. 
In the EMRI systems we consider here, the behavior of the DF is understood in a similar manner in Fig.~\ref{fig:phase_space}, as the central Kerr BH impacts the proportion of prograde/retrograde particles encountered by the secondary body. 
Given the primary spin and the halo angular momentum lead to qualitatively similar effects for the DF loss, we wish to understand whether damping and sign reversal in the DF force seen in Fig.~\ref{fig:DF_acc_Kerr} may be mimicked for non-spinning primaries by adding sufficient angular momentum dependence in the initial DM environment.

In Newtonian computations, the distribution function describing an ensemble of gravitating particles is not uniquely determined from the density. Indeed, we may introduce a net motion about the $z$-axis which changes higher moments of the distribution while preserving the density profile. Inspired by the Newtonian case, let us impart an $L_z$ dependence into the Hernquist distribution function by taking 
\begin{equation}
\label{eq:angular_distribution_function}
f(E,L_z)= f(E)\left[1+\kappa\tanh\left(\frac{L_z}{L_0}\right)\right],
\end{equation}
where $-1\leq\kappa\leq1$, and $L_0$ is an angular momentum scale. Positive (negative) values of $\kappa$ give higher weight to positive (negative) values of $L_z$. The scale $L_0$ controls how fast the $L_z$ weighting ``turns on'' near low values of $L_z$. For large $L_0$, $f(E,L_z)\sim f(E)(1+\kappa L_z/L_0)$, such that the $L_z$ dependence is quite mild. As $L_0\rightarrow0$, the distribution function becomes $f(E,L_z)\sim f(E)[1-\kappa+2\kappa\Theta(L_z)]$, where $\Theta(L_z)$ is the Heaviside step function. 
In the relativistic adiabatic growth models discussed in Sec.~\ref{sec:adiabatic_growth}, the matching of angular invariants $I'_\phi(L_z)=I_\phi(\mathcal{L}_z)$ predicts $\tanh(L_z/L_0)=\tanh(\mathcal{L}_z/L_0)$ for an axially symmetric background. In Schwarzschild, the limits on $\mathcal{L}_z$ due to the loss cone conditions are symmetric, such that $\mathcal{L}_z^{\text{max}}=-\mathcal{L}^{\text{min}}_z=\mathcal{C}^{1/2}\sin\theta$. When integrating over the allowed phase space of orbital parameters to obtain the density distribution, contributions from the $\tanh(\mathcal{L}_z/L_0)$ term vanish due to the $\mathcal{L}_z$ symmetry, and we obtain the same $\rho(r)$ regardless of the choice of $(\kappa,\,L_0)$ as in the Newtonian case. This argument holds for any choice of odd function of $L_z$, 
but does not hold in the Kerr background, as the allowed phase space region in $\mathcal{L}_z$ is no longer symmetric (see, e.g., Fig. 6 of Ref.~\cite{Ferrer:2017xwm}).

\begin{figure}[t]
\centering 
\includegraphics[width=\linewidth]{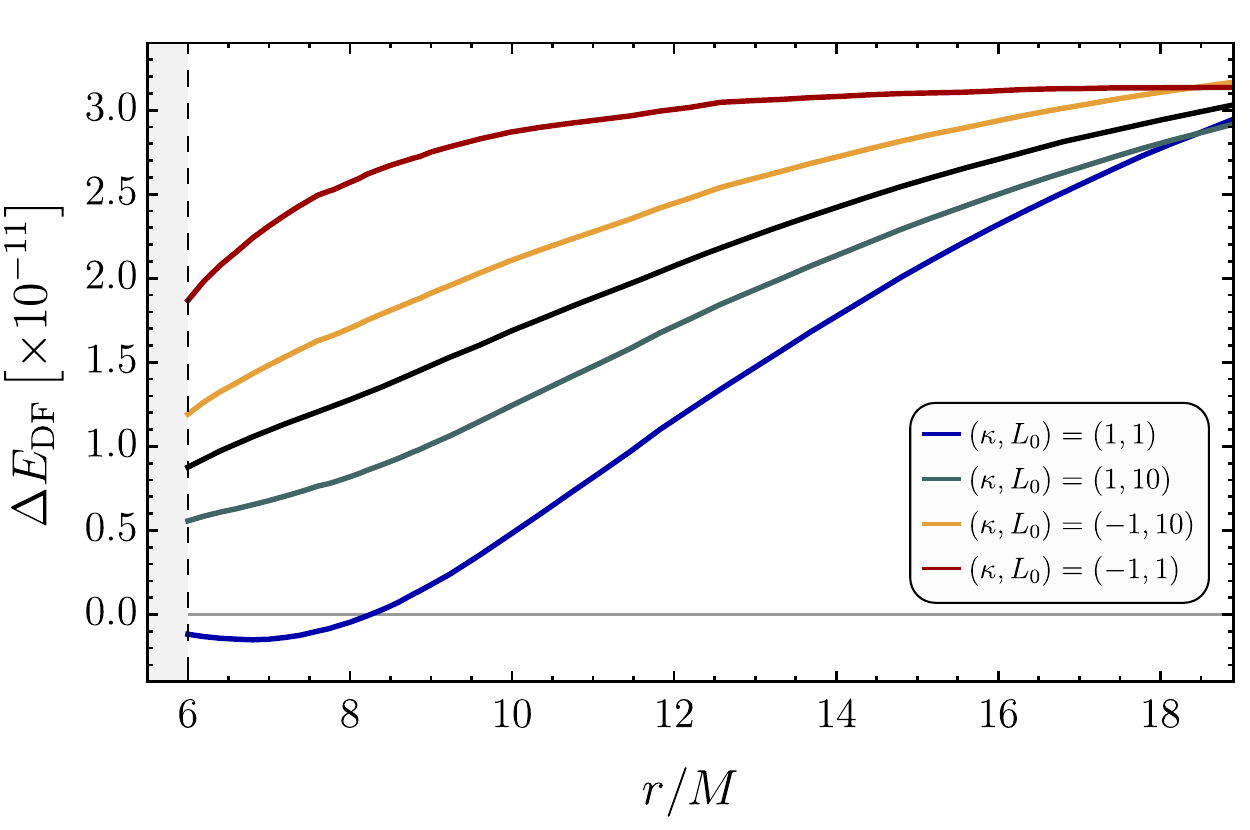}
\caption{DF energy loss through one orbital cycle in a Schwarzschild background, computed via Eq.~\eqref{eq:acceleration_DF_ACC} and adopting an initial $L_z$ dependence in the distribution function given by Eq.~\eqref{eq:angular_distribution_function}. For large enough positive $\kappa$ and small enough $L_0$, the sign reversal in the friction can be mimicked without the need for frame dragging. Alternatively, friction can also be enhanced by taking $\kappa$ negative, such that a greater percentage of particles move with negative $v^3$ in the secondary body's frame. The black line is the acceleration assuming no initial anisotropy in the distribution function ($\kappa=0$). We choose the halo mass, scale radius, central BH mass, and secondary mass to be the same as Fig.~\ref{fig:DF_acc_Kerr}.}
\label{fig:friction_angular_dist_function}
\end{figure}

While the form of the density distribution is preserved, the accelerations due to scattering and accretion onto the secondary are significantly impacted by the angular momentum dependence in the DM distribution. In Fig.~\ref{fig:friction_angular_dist_function}, we compute the energy loss due to DF in a Schwarzschild background using the $L_z$-dependent distribution function of Eq.~\eqref{eq:angular_distribution_function}, for various choices of the parameters $(\kappa,L_0)$. In the secondary body's frame, the shape of the loss cone remains identical to the leftmost column of Fig.~\ref{fig:phase_space}, but the weighting of positive/negative $v^3$ contributions to the phase space is altered. Positive (negative) values of $\kappa$ control whether the weighting skews towards positive (negative) $v^3$ values, and $L_0$ controls how quickly the weighting ``turns on'' at small values of $L_z$. For sufficiently large values of $\kappa$ and sufficiently small values of $L_0$, we find that the sign reversal of the friction seen in the Kerr case may be mimicked by introducing initial rotation into the DM halo. Alternately, taking negative $\kappa$ leads to an enhancement of the frictional effects, as a greater number of particles are moving with negative $L_z$ ``against'' the motion of the secondary. This is to be contrasted with Fig.~\ref{fig:DF_acc_Kerr}, in which the sign reversal happens solely as a consequence of the background spacetime's frame dragging forcing the loss cone to favor $v^3>0$ particles, regardless of the anisotropy of the initial distribution function. In that case, frictional effects may still be enhanced or weakened as a consequence of the distribution function's initial anisotropy, and a further systematic exploration of these effects and subsequent imprint on EMRI orbital dynamics is warranted. These results further highlight the importance of the more complete friction prescription given by Eq.~\eqref{eq:acceleration_DF_ACC} when treating EMRIs evolving in DM environments, as prescriptions of the DF force which only depend on local densities would be unable to capture the detailed behavior shown in Fig.~\ref{fig:friction_angular_dist_function}.

While the damping and sign reversal of the friction may be mimicked by adopting an $L_z$ dependence in the distribution function, caution must be taken such that the initial anisotropy imposed by the choice of $(\kappa,L_0)$ is physically realistic. The model in Eq.~\eqref{eq:angular_distribution_function} best matches cosmological simulations of DM halos for very low $L_0$ values~\cite{Chiba:2023rqm}, but severely underestimates the spherical average $\langle L_z\rangle$ in the inner galactic region. Further, choices of $(|\kappa|,L_0)=(1,1)$ correspond to highly anisotropic initial halos with a significant fraction of particles moving coherently in the $\text{sgn}(\kappa)\boldsymbol{\pd}_{\varphi}$ direction, which is unlikely to occur naturally. Models with dependence on both $L$ and $L_z$ perform much better without the need for significant anisotropy in the DM distribution. As such, the results in this section should be taken as a proof of principle that the damping and sign-flipping in the DF force may occur solely due to the chosen distribution function. These effects are more likely to occur as a consequence of the primary BH's spin, however, as initial halos without any $L_z$ dependence can have significant anisotropy in their final $\mathcal{L}_z$ distribution in the region where frame-dragging becomes important. It is unclear whether the similarity of frame-dragging induced anisotropy and global rotation in the DM halo would present issues for studies of DM with EMRI GW signals. A comprehensive exploration of the potential degeneracy of these effects and their impact on GW data analysis warrants future investigation.
\subsection{Comparison with Chandrasekhar-like prescriptions}
\label{sec:comparison}

As pointed out in Ref.~\cite{Vicente:2025gsg} and in Sec.~\ref{sec:DF_acc_in_Kerr}, prescriptions of DF which rely solely on the value of the local density and on the perturber's velocity are insufficient for EMRI systems, since particle distributions in the relativistic regime are generally anisotropic. Many prior works have relied on results from linear motion through homogeneous isotropic media~\cite{Speeney:2022ryg,Mitra:2025tag,Traykova:2021dua,Traykova:2023qyv}, and have adopted special relativistic correction factors for the friction force. Contrary to claims in Ref.~\cite{Vicente:2025gsg}, such prescriptions may actually over-estimate the importance of DF, as we show below.

Here we compare several prescriptions for the DF force and demonstrate how and why the classical results differ from the relativistic prescription of Eq.~\eqref{eq:acceleration_DF_ACC}. We primarily want to understand whether or not computations may be simplified through some convenient approximations, or if the full integration over the phase space is always necessary. We consider two models as points of comparison: (i) the classical Chandrasekhar formulation of DF for a body moving linearly in an isotropic homogeneous medium~\cite{1943ApJ....97..255C,Kavanagh:2024lgq}; (ii) the short wavelength particle limit of DF induced by massive scalar fields, derived from relativistic scattering in BH backgrounds~\cite{Traykova:2021dua,Traykova:2023qyv,Vicente:2022ivh}. These models may be described via the expression
\begin{equation}
\label{eq:Chandra_Traykova_DF}
\mathcal{A}^{\text{Ch/Tr}}=\frac{4\pi}{u^2}\zeta(u)\rho(r)\xi_{|v|<u}\Lambda^{\text{Ch/Tr}},
\end{equation}
where $\rho(r)$ is the local density distribution at the location of the moving body, $\xi_{|v|<u}$ is the fraction of particles moving with velocities less than the secondary, $\Lambda^{\text{Ch/Tr}}$ is the Coulomb logarithm in the Chandrasekhar and Traykova prescriptions, respectively (see below), and $u$ is the magnitude of the secondary's velocity. The factor $\zeta(u)=\gamma^2_u(1+u^2)^2=(1+u^2)^2/(1-u^2)$ incorporates special relativistic corrections, which appear as a consequence of length contraction and scattering angle modifications in the body's frame. In Refs.~\cite{Speeney:2022ryg,Mitra:2025tag}, the Coulomb logarithm was set to $\ln\Lambda\sim3$, which produced results quantitatively similar to the fully relativistic computation. Here, we instead adopt physically motivated expressions for the Coulomb logarithm, while setting the maximum impact parameter to be the Hill radius for consistency with our relativistic results. In the Chandrasekhar-like expression, we take $\Lambda^{\text{Ch}}\approx\ln\sqrt{1+r^2u^4/(3q^{2})^{2/3}}$~\cite{Kavanagh:2024lgq}, and remove the $\zeta(u)$ factor. For the massive scalar case, we keep the $\zeta(u)$ factor and, following Traykova et al.~\cite{Traykova:2023qyv}, we take the Coulomb logarithm to be
\begin{equation}
\label{eq:Traykova_DF_Log}
\begin{aligned}
\Lambda^{\text{Tr}}(u)&=\ln\left[\frac{ru/(3q^2)^{1/3}}{10(1+u^2)(b_{\text{cr}}/m)}\right]\\
&~~~+3.7+9.24u^2-12.16u^3+4.79u^4,
\end{aligned}
\end{equation}
where we have again set the maximum impact parameter to the Hill radius for a consistent comparison. 

\begin{figure}
    \centering
    \includegraphics[width=\linewidth]{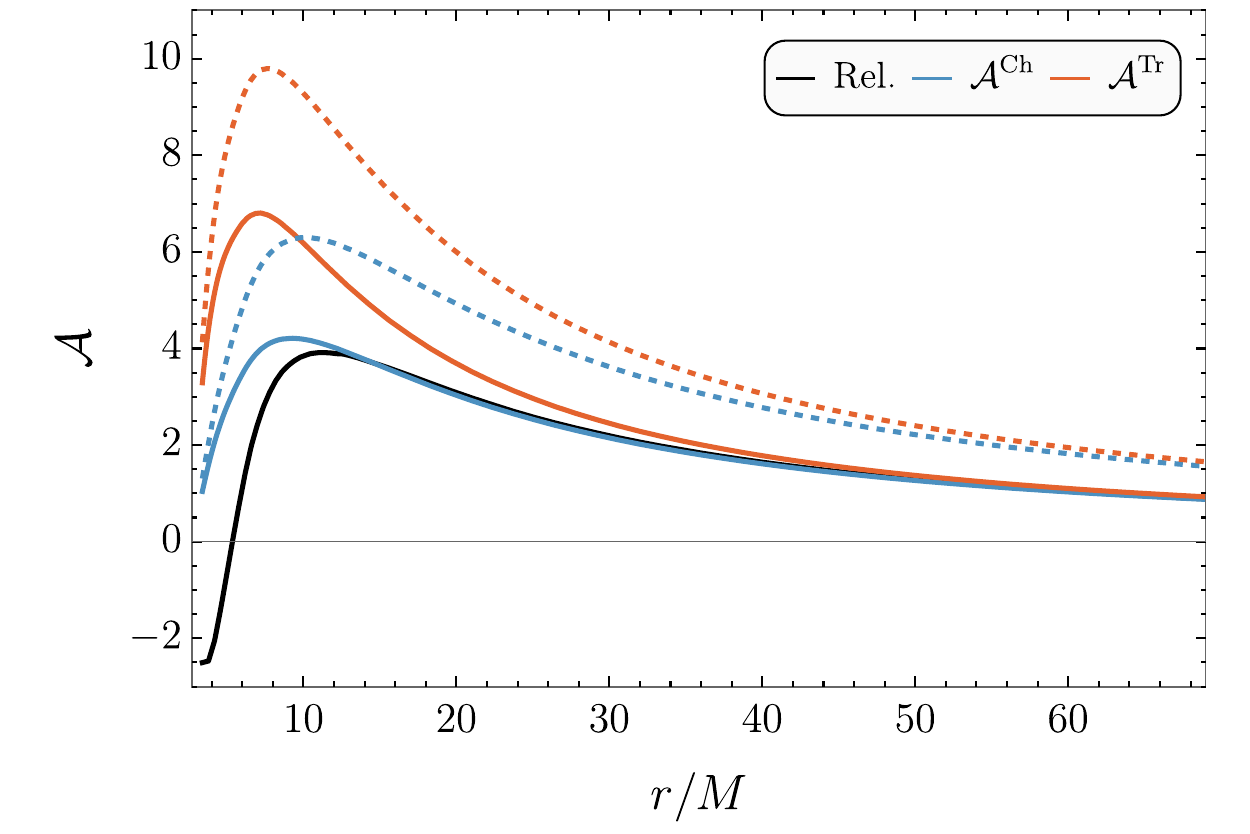}
    \caption{Comparison of the relativistic DF acceleration computed via Eq.~\eqref{eq:acceleration_DF_ACC} to simpler prescriptions which assume an infinite homogeneous medium. We take the same halo and binary configuration as in Fig.~\ref{fig:DF_acc_Kerr}, and $a=0.7$. The black line is the relativistic DF, and the other solid lines are computed using Eq.~\eqref{eq:Chandra_Traykova_DF}. The dashed lines are the same classical expressions with the $\xi_{|v|<u}$ factor removed.  
    }
    \label{fig:DF_comparison}
\end{figure}

In Fig.~\ref{fig:DF_comparison}, we compare the relativistic acceleration computed via Eq.~\eqref{eq:acceleration_DF_ACC} to the DF prescriptions in Eq.~\eqref{eq:Chandra_Traykova_DF}. We take the secondary's velocity to be its circular speed in the ZAMO frame, $u_{\text{ZAMO}}=\Omega\Delta^{-1/2}(r^2-2a\sqrt{r}+a^2)$, and use the adiabatically grown DM density as perceived in that frame. To estimate the factor $\xi_{|v|<u}$, we assume that particles are evenly distributed in positive and negative velocities, in line with the assumptions of infinite, homogeneous media. With this assumption, the fraction of particles moving more slowly than the secondary in the ZAMO frame which contribute to the DF may be estimated as $\xi_{|v|<u}=(1+|u_{\text{ZAMO}}|)/2$. This choice is not generally applicable near the primary, since the distribution function is highly anisotropic in that region due to the loss cone conditions. Nevertheless, we wish to highlight how such choices (which have been used in studies on the impact of DM-induced friction in EMRI systems) are insufficient. In the weak-field regime, the fully relativistic friction results in Fig.~\ref{fig:DF_comparison} match quite closely to the isotropic cases. Friction results start to deviate below $r/M\sim20$, and the sign reversal in the DF seen in the relativistic curve is not adequately captured by the simpler prescriptions. As such, DM-induced effects on EMRI GW signals may be significantly overestimated if one adopts the wrong DF prescription, as the friction damping seen in highly spinning primaries goes unaccounted for. If the $\xi_{|v|<u}$ factor is ignored entirely in Eq.~\eqref{eq:Chandra_Traykova_DF}, the agreement with the isotropic case becomes much worse. 

\section{Conclusions}
\label{sec:conclusions}

We have extended the relativistic framework of Ref.~\cite{Vicente:2025gsg} for DF and accretion in collisionless media to a Kerr background, considering a non-spinning secondary on circular equatorial orbits around a spinning primary embedded in an adiabatically grown DM spike. The friction and accretion follow from integrating relativistic scattering and capture cross sections over the DM phase space as seen in the secondary's rest frame, with loss-cone conditions that guarantee only bound, non-plunging particles contribute. Even for an initially isotropic halo, this phase space is anisotropic, both because of the secondary's motion and because of the structure of the loss cone in Kerr.

At large orbital radii, DF approaches the Newtonian expectation and scales with the local density, which increases with the primary spin. Closer to the BH, the behavior is qualitatively different. For $a\gtrsim0.37$ the prograde ISCO lies inside the retrograde marginally bound radius $r^-_{\rm mb}$, so the secondary can orbit in a region where frame dragging forces all the bound DM particles to co-rotate. As a consequence, particles overtaking the secondary dominate the momentum transfer: the friction weakens and eventually changes sign, even though the DM density reaches its maximum in this region. Accretion remains an order of magnitude smaller than scattering/DF and does not change sign for the configurations we have considered. This behavior cannot be captured by any prescription in which the friction is proportional to the local density.

We also showed that the damping and reversal of the friction can be reproduced around a Schwarzschild BH by endowing the initial halo with net angular momentum through an $\mathcal{L}_z$-dependent distribution function, while negative net rotation enhances the friction. Reproducing the Kerr behavior in this way requires anisotropies much larger than those found in cosmological simulations, so the effect is far more naturally attributed to the spin of the central BH. Nevertheless, these results show that the friction is sensitive not only to the density profile, but also to the detailed velocity structure of the environment. Isotropic Chandrasekhar-like prescriptions miss this structure and can overestimate the DF by a factor of a few near its peak.

Several extensions are needed before these effects can be included in precision EMRI waveforms. The location of the sign reversal depends on the maximum impact parameter, which we set to the Hill radius; self-consistent choices may follow from effective field theory treatments~\cite{Modrekiladze:2026twz,Wilson-Gerow:2026plu,Mitra:2026gar}. The response also depends on the nature of the secondary, and neutron stars, through which DM can pass, may behave differently from BHs~\cite{Karydas:2025bkj,Karydas:2025gqp}. For eccentric or inclined orbits the accelerations acquire radial and polar components, and the evolution of the Carter constant becomes nontrivial. Finally, we have neglected the back-reaction of the secondary on the DM distribution, which may be important for intermediate-mass-ratio systems, as well as the gravitational coupling between the environment and the emitted radiation. Combining the accelerations computed here with radiation-reaction fluxes to evolve inspirals and assess the resulting dephasing for LISA is an important topic for future work.

\noindent
{\bf \textit{Note added.}} During the preparation of this manuscript, a similar work producing much of the same results appeared~\cite{Xiong:2026fqn}. We have cross-checked, and the results presented in this work are consistent with those in Ref.~\cite{Xiong:2026fqn}.

\begin{acknowledgments}
We thank Rodrigo Vicente for stimulating discussions on this work. 
N.S. and E.B. are supported by NSF Grants No.~AST-2606672, No.~PHY-2513337, No.~PHY-090003, and No.~PHY-20043, by the Simons Foundation [MPS-SIP-00001698, E.B.], and by the Simons Foundation International [SFI-MPS-BH-00012593-02]. This project has received funding from the European Union’s Horizon Europe research and innovation programme under the Marie Skłodowska-Curie Staﬀ Exchanges grant agreement No.~101299389 (STRONG).
This work was carried out at the Advanced Research Computing at Hopkins (ARCH) core facility (\url{https://www.arch.jhu.edu/}), which is supported by the NSF Grant No. OAC-1920103. The research of S.C. is supported by MATRICS (MTR/2023/000049) and Core Research Grants (CRG/2023/000934) from SERB, ANRF, Government of India.
\end{acknowledgments}

\bibliography{refs}

\begin{thebibliography}{80}%
\makeatletter
\providecommand \@ifxundefined [1]{%
 \@ifx{#1\undefined}
}%
\providecommand \@ifnum [1]{%
 \ifnum #1\expandafter \@firstoftwo
 \else \expandafter \@secondoftwo
 \fi
}%
\providecommand \@ifx [1]{%
 \ifx #1\expandafter \@firstoftwo
 \else \expandafter \@secondoftwo
 \fi
}%
\providecommand \natexlab [1]{#1}%
\providecommand \enquote  [1]{``#1''}%
\providecommand \bibnamefont  [1]{#1}%
\providecommand \bibfnamefont [1]{#1}%
\providecommand \citenamefont [1]{#1}%
\providecommand \href@noop [0]{\@secondoftwo}%
\providecommand \href [0]{\begingroup \@sanitize@url \@href}%
\providecommand \@href[1]{\@@startlink{#1}\@@href}%
\providecommand \@@href[1]{\endgroup#1\@@endlink}%
\providecommand \@sanitize@url [0]{\catcode `\\12\catcode `\$12\catcode
  `\&12\catcode `\#12\catcode `\^12\catcode `\_12\catcode `\%12\relax}%
\providecommand \@@startlink[1]{}%
\providecommand \@@endlink[0]{}%
\providecommand \url  [0]{\begingroup\@sanitize@url \@url }%
\providecommand \@url [1]{\endgroup\@href {#1}{\urlprefix }}%
\providecommand \urlprefix  [0]{URL }%
\providecommand \Eprint [0]{\href }%
\providecommand \doibase [0]{http://dx.doi.org/}%
\providecommand \selectlanguage [0]{\@gobble}%
\providecommand \bibinfo  [0]{\@secondoftwo}%
\providecommand \bibfield  [0]{\@secondoftwo}%
\providecommand \translation [1]{[#1]}%
\providecommand \BibitemOpen [0]{}%
\providecommand \bibitemStop [0]{}%
\providecommand \bibitemNoStop [0]{.\EOS\space}%
\providecommand \EOS [0]{\spacefactor3000\relax}%
\providecommand \BibitemShut  [1]{\csname bibitem#1\endcsname}%
\let\auto@bib@innerbib\@empty
%</preamble>
\bibitem [{\citenamefont {Freese}(2009)}]{Freese:2008cz}%
  \BibitemOpen
  \bibfield  {author} {\bibinfo {author} {\bibfnamefont {K.}~\bibnamefont
  {Freese}},\ }\href {\doibase 10.1051/eas/0936016} {\bibfield  {journal}
  {\bibinfo  {journal} {EAS Publ. Ser.}\ }\textbf {\bibinfo {volume} {36}},\
  \bibinfo {pages} {113} (\bibinfo {year} {2009})},\ \Eprint
  {http://arxiv.org/abs/0812.4005} {arXiv:0812.4005 [astro-ph]} \BibitemShut
  {NoStop}%
\bibitem [{\citenamefont {Aghanim}\ \emph {et~al.}(2020)\citenamefont {Aghanim}
  \emph {et~al.}}]{Planck:2018vyg}%
  \BibitemOpen
  \bibfield  {author} {\bibinfo {author} {\bibfnamefont {N.}~\bibnamefont
  {Aghanim}} \emph {et~al.} (\bibinfo {collaboration} {Planck}),\ }\href
  {\doibase 10.1051/0004-6361/201833910} {\bibfield  {journal} {\bibinfo
  {journal} {Astron. Astrophys.}\ }\textbf {\bibinfo {volume} {641}},\ \bibinfo
  {pages} {A6} (\bibinfo {year} {2020})},\ \bibinfo {note} {[Erratum:
  Astron.Astrophys. 652, C4 (2021)]},\ \Eprint
  {http://arxiv.org/abs/1807.06209} {arXiv:1807.06209 [astro-ph.CO]}
  \BibitemShut {NoStop}%
\bibitem [{\citenamefont {Bullock}\ and\ \citenamefont
  {Boylan-Kolchin}(2017)}]{Bullock:2017xww}%
  \BibitemOpen
  \bibfield  {author} {\bibinfo {author} {\bibfnamefont {J.~S.}\ \bibnamefont
  {Bullock}}\ and\ \bibinfo {author} {\bibfnamefont {M.}~\bibnamefont
  {Boylan-Kolchin}},\ }\href {\doibase 10.1146/annurev-astro-091916-055313}
  {\bibfield  {journal} {\bibinfo  {journal} {Annu. Rev. Astron. Astrophys.}\
  }\textbf {\bibinfo {volume} {55}},\ \bibinfo {pages} {343} (\bibinfo {year}
  {2017})},\ \Eprint {http://arxiv.org/abs/1707.04256} {arXiv:1707.04256
  [astro-ph.CO]} \BibitemShut {NoStop}%
\bibitem [{\citenamefont {Flores}\ and\ \citenamefont
  {Primack}(1994)}]{Flores:1994gz}%
  \BibitemOpen
  \bibfield  {author} {\bibinfo {author} {\bibfnamefont {R.~A.}\ \bibnamefont
  {Flores}}\ and\ \bibinfo {author} {\bibfnamefont {J.~R.}\ \bibnamefont
  {Primack}},\ }\href {\doibase 10.1086/187350} {\bibfield  {journal} {\bibinfo
   {journal} {Astrophys. J. Lett.}\ }\textbf {\bibinfo {volume} {427}},\
  \bibinfo {pages} {L1} (\bibinfo {year} {1994})},\ \Eprint
  {http://arxiv.org/abs/astro-ph/9402004} {arXiv:astro-ph/9402004} \BibitemShut
  {NoStop}%
\bibitem [{\citenamefont {de~Blok}(2010)}]{deBlok:2009sp}%
  \BibitemOpen
  \bibfield  {author} {\bibinfo {author} {\bibfnamefont {W.~J.~G.}\
  \bibnamefont {de~Blok}},\ }\href {\doibase 10.1155/2010/789293} {\bibfield
  {journal} {\bibinfo  {journal} {Adv. Astron.}\ }\textbf {\bibinfo {volume}
  {2010}},\ \bibinfo {pages} {789293} (\bibinfo {year} {2010})},\ \Eprint
  {http://arxiv.org/abs/0910.3538} {arXiv:0910.3538 [astro-ph.CO]} \BibitemShut
  {NoStop}%
\bibitem [{\citenamefont {Klypin}\ \emph {et~al.}(1999)\citenamefont {Klypin},
  \citenamefont {Kravtsov}, \citenamefont {Valenzuela},\ and\ \citenamefont
  {Prada}}]{Klypin:1999uc}%
  \BibitemOpen
  \bibfield  {author} {\bibinfo {author} {\bibfnamefont {A.~A.}\ \bibnamefont
  {Klypin}}, \bibinfo {author} {\bibfnamefont {A.~V.}\ \bibnamefont
  {Kravtsov}}, \bibinfo {author} {\bibfnamefont {O.}~\bibnamefont
  {Valenzuela}}, \ and\ \bibinfo {author} {\bibfnamefont {F.}~\bibnamefont
  {Prada}},\ }\href {\doibase 10.1086/307643} {\bibfield  {journal} {\bibinfo
  {journal} {Astrophys. J.}\ }\textbf {\bibinfo {volume} {522}},\ \bibinfo
  {pages} {82} (\bibinfo {year} {1999})},\ \Eprint
  {http://arxiv.org/abs/astro-ph/9901240} {arXiv:astro-ph/9901240} \BibitemShut
  {NoStop}%
\bibitem [{\citenamefont {Moore}\ \emph {et~al.}(1999)\citenamefont {Moore},
  \citenamefont {Ghigna}, \citenamefont {Governato}, \citenamefont {Lake},
  \citenamefont {Quinn}, \citenamefont {Stadel},\ and\ \citenamefont
  {Tozzi}}]{Moore:1999nt}%
  \BibitemOpen
  \bibfield  {author} {\bibinfo {author} {\bibfnamefont {B.}~\bibnamefont
  {Moore}}, \bibinfo {author} {\bibfnamefont {S.}~\bibnamefont {Ghigna}},
  \bibinfo {author} {\bibfnamefont {F.}~\bibnamefont {Governato}}, \bibinfo
  {author} {\bibfnamefont {G.}~\bibnamefont {Lake}}, \bibinfo {author}
  {\bibfnamefont {T.~R.}\ \bibnamefont {Quinn}}, \bibinfo {author}
  {\bibfnamefont {J.}~\bibnamefont {Stadel}}, \ and\ \bibinfo {author}
  {\bibfnamefont {P.}~\bibnamefont {Tozzi}},\ }\href {\doibase 10.1086/312287}
  {\bibfield  {journal} {\bibinfo  {journal} {Astrophys. J. Lett.}\ }\textbf
  {\bibinfo {volume} {524}},\ \bibinfo {pages} {L19} (\bibinfo {year}
  {1999})},\ \Eprint {http://arxiv.org/abs/astro-ph/9907411}
  {arXiv:astro-ph/9907411} \BibitemShut {NoStop}%
\bibitem [{\citenamefont {Bertone}\ \emph {et~al.}(2005)\citenamefont
  {Bertone}, \citenamefont {Hooper},\ and\ \citenamefont
  {Silk}}]{Bertone:2004pz}%
  \BibitemOpen
  \bibfield  {author} {\bibinfo {author} {\bibfnamefont {G.}~\bibnamefont
  {Bertone}}, \bibinfo {author} {\bibfnamefont {D.}~\bibnamefont {Hooper}}, \
  and\ \bibinfo {author} {\bibfnamefont {J.}~\bibnamefont {Silk}},\ }\href
  {\doibase 10.1016/j.physrep.2004.08.031} {\bibfield  {journal} {\bibinfo
  {journal} {Phys. Rep.}\ }\textbf {\bibinfo {volume} {405}},\ \bibinfo {pages}
  {279} (\bibinfo {year} {2005})},\ \Eprint
  {http://arxiv.org/abs/hep-ph/0404175} {arXiv:hep-ph/0404175} \BibitemShut
  {NoStop}%
\bibitem [{\citenamefont {Cirelli}\ \emph {et~al.}(2024)\citenamefont
  {Cirelli}, \citenamefont {Strumia},\ and\ \citenamefont
  {Zupan}}]{Cirelli:2024ssz}%
  \BibitemOpen
  \bibfield  {author} {\bibinfo {author} {\bibfnamefont {M.}~\bibnamefont
  {Cirelli}}, \bibinfo {author} {\bibfnamefont {A.}~\bibnamefont {Strumia}}, \
  and\ \bibinfo {author} {\bibfnamefont {J.}~\bibnamefont {Zupan}},\ }\href
  {\doibase 10.21468/SciPostPhysRev.1} {\  (\bibinfo {year} {2024}),\
  10.21468/SciPostPhysRev.1},\ \Eprint {http://arxiv.org/abs/2406.01705}
  {arXiv:2406.01705 [hep-ph]} \BibitemShut {NoStop}%
\bibitem [{\citenamefont {Amaro-Seoane}\ \emph {et~al.}(2017)\citenamefont
  {Amaro-Seoane} \emph {et~al.}}]{LISA:2017pwj}%
  \BibitemOpen
  \bibfield  {author} {\bibinfo {author} {\bibfnamefont {P.}~\bibnamefont
  {Amaro-Seoane}} \emph {et~al.} (\bibinfo {collaboration} {LISA}),\
  }\href@noop {} {\  (\bibinfo {year} {2017})},\ \Eprint
  {http://arxiv.org/abs/1702.00786} {arXiv:1702.00786 [astro-ph.IM]}
  \BibitemShut {NoStop}%
\bibitem [{\citenamefont {Seoane}\ \emph {et~al.}(2023)\citenamefont {Seoane}
  \emph {et~al.}}]{LISA:2022yao}%
  \BibitemOpen
  \bibfield  {author} {\bibinfo {author} {\bibfnamefont {P.~A.}\ \bibnamefont
  {Seoane}} \emph {et~al.} (\bibinfo {collaboration} {LISA}),\ }\href {\doibase
  10.1007/s41114-022-00041-y} {\bibfield  {journal} {\bibinfo  {journal}
  {Living Rev. Relativ.}\ }\textbf {\bibinfo {volume} {26}},\ \bibinfo {pages}
  {2} (\bibinfo {year} {2023})},\ \Eprint {http://arxiv.org/abs/2203.06016}
  {arXiv:2203.06016 [gr-qc]} \BibitemShut {NoStop}%
\bibitem [{\citenamefont {Colpi}\ \emph {et~al.}(2024)\citenamefont {Colpi}
  \emph {et~al.}}]{LISA:2024hlh}%
  \BibitemOpen
  \bibfield  {author} {\bibinfo {author} {\bibfnamefont {M.}~\bibnamefont
  {Colpi}} \emph {et~al.} (\bibinfo {collaboration} {LISA}),\ }\href@noop {} {\
   (\bibinfo {year} {2024})},\ \Eprint {http://arxiv.org/abs/2402.07571}
  {arXiv:2402.07571 [astro-ph.CO]} \BibitemShut {NoStop}%
\bibitem [{\citenamefont {Cardoso}\ \emph
  {et~al.}(2022{\natexlab{a}})\citenamefont {Cardoso}, \citenamefont
  {Destounis}, \citenamefont {Duque}, \citenamefont {Macedo},\ and\
  \citenamefont {Maselli}}]{Cardoso:2021wlq}%
  \BibitemOpen
  \bibfield  {author} {\bibinfo {author} {\bibfnamefont {V.}~\bibnamefont
  {Cardoso}}, \bibinfo {author} {\bibfnamefont {K.}~\bibnamefont {Destounis}},
  \bibinfo {author} {\bibfnamefont {F.}~\bibnamefont {Duque}}, \bibinfo
  {author} {\bibfnamefont {R.~P.}\ \bibnamefont {Macedo}}, \ and\ \bibinfo
  {author} {\bibfnamefont {A.}~\bibnamefont {Maselli}},\ }\href {\doibase
  10.1103/PhysRevD.105.L061501} {\bibfield  {journal} {\bibinfo  {journal}
  {Phys. Rev. D}\ }\textbf {\bibinfo {volume} {105}},\ \bibinfo {pages}
  {L061501} (\bibinfo {year} {2022}{\natexlab{a}})},\ \Eprint
  {http://arxiv.org/abs/2109.00005} {arXiv:2109.00005 [gr-qc]} \BibitemShut
  {NoStop}%
\bibitem [{\citenamefont {Cardoso}\ \emph
  {et~al.}(2022{\natexlab{b}})\citenamefont {Cardoso}, \citenamefont
  {Destounis}, \citenamefont {Duque}, \citenamefont {Panosso~Macedo},\ and\
  \citenamefont {Maselli}}]{Cardoso:2022whc}%
  \BibitemOpen
  \bibfield  {author} {\bibinfo {author} {\bibfnamefont {V.}~\bibnamefont
  {Cardoso}}, \bibinfo {author} {\bibfnamefont {K.}~\bibnamefont {Destounis}},
  \bibinfo {author} {\bibfnamefont {F.}~\bibnamefont {Duque}}, \bibinfo
  {author} {\bibfnamefont {R.}~\bibnamefont {Panosso~Macedo}}, \ and\ \bibinfo
  {author} {\bibfnamefont {A.}~\bibnamefont {Maselli}},\ }\href {\doibase
  10.1103/PhysRevLett.129.241103} {\bibfield  {journal} {\bibinfo  {journal}
  {Phys. Rev. Lett.}\ }\textbf {\bibinfo {volume} {129}},\ \bibinfo {pages}
  {241103} (\bibinfo {year} {2022}{\natexlab{b}})},\ \Eprint
  {http://arxiv.org/abs/2210.01133} {arXiv:2210.01133 [gr-qc]} \BibitemShut
  {NoStop}%
\bibitem [{\citenamefont {Figueiredo}\ \emph {et~al.}(2023)\citenamefont
  {Figueiredo}, \citenamefont {Maselli},\ and\ \citenamefont
  {Cardoso}}]{Figueiredo:2023gas}%
  \BibitemOpen
  \bibfield  {author} {\bibinfo {author} {\bibfnamefont {E.}~\bibnamefont
  {Figueiredo}}, \bibinfo {author} {\bibfnamefont {A.}~\bibnamefont {Maselli}},
  \ and\ \bibinfo {author} {\bibfnamefont {V.}~\bibnamefont {Cardoso}},\ }\href
  {\doibase 10.1103/PhysRevD.107.104033} {\bibfield  {journal} {\bibinfo
  {journal} {Phys. Rev. D}\ }\textbf {\bibinfo {volume} {107}},\ \bibinfo
  {pages} {104033} (\bibinfo {year} {2023})},\ \Eprint
  {http://arxiv.org/abs/2303.08183} {arXiv:2303.08183 [gr-qc]} \BibitemShut
  {NoStop}%
\bibitem [{\citenamefont {Gliorio}\ \emph {et~al.}(2025)\citenamefont
  {Gliorio}, \citenamefont {Berti}, \citenamefont {Maselli},\ and\
  \citenamefont {Speeney}}]{Gliorio:2025cbh}%
  \BibitemOpen
  \bibfield  {author} {\bibinfo {author} {\bibfnamefont {S.}~\bibnamefont
  {Gliorio}}, \bibinfo {author} {\bibfnamefont {E.}~\bibnamefont {Berti}},
  \bibinfo {author} {\bibfnamefont {A.}~\bibnamefont {Maselli}}, \ and\
  \bibinfo {author} {\bibfnamefont {N.}~\bibnamefont {Speeney}},\ }\href
  {\doibase 10.1103/dw6c-14pt} {\bibfield  {journal} {\bibinfo  {journal}
  {Phys. Rev. D}\ }\textbf {\bibinfo {volume} {112}},\ \bibinfo {pages}
  {124050} (\bibinfo {year} {2025})},\ \Eprint
  {http://arxiv.org/abs/2503.16649} {arXiv:2503.16649 [gr-qc]} \BibitemShut
  {NoStop}%
\bibitem [{\citenamefont {Speeney}\ \emph {et~al.}(2024)\citenamefont
  {Speeney}, \citenamefont {Berti}, \citenamefont {Cardoso},\ and\
  \citenamefont {Maselli}}]{Speeney:2024mas}%
  \BibitemOpen
  \bibfield  {author} {\bibinfo {author} {\bibfnamefont {N.}~\bibnamefont
  {Speeney}}, \bibinfo {author} {\bibfnamefont {E.}~\bibnamefont {Berti}},
  \bibinfo {author} {\bibfnamefont {V.}~\bibnamefont {Cardoso}}, \ and\
  \bibinfo {author} {\bibfnamefont {A.}~\bibnamefont {Maselli}},\ }\href
  {\doibase 10.1103/PhysRevD.109.084068} {\bibfield  {journal} {\bibinfo
  {journal} {Phys. Rev. D}\ }\textbf {\bibinfo {volume} {109}},\ \bibinfo
  {pages} {084068} (\bibinfo {year} {2024})},\ \Eprint
  {http://arxiv.org/abs/2401.00932} {arXiv:2401.00932 [gr-qc]} \BibitemShut
  {NoStop}%
\bibitem [{\citenamefont {Xu}\ \emph {et~al.}(2026{\natexlab{a}})\citenamefont
  {Xu}, \citenamefont {Brito}, \citenamefont {Della~Monica}, \citenamefont
  {Vicente},\ and\ \citenamefont {Yuan}}]{Xu:2026aic}%
  \BibitemOpen
  \bibfield  {author} {\bibinfo {author} {\bibfnamefont {Q.-X.}\ \bibnamefont
  {Xu}}, \bibinfo {author} {\bibfnamefont {R.}~\bibnamefont {Brito}}, \bibinfo
  {author} {\bibfnamefont {R.}~\bibnamefont {Della~Monica}}, \bibinfo {author}
  {\bibfnamefont {R.}~\bibnamefont {Vicente}}, \ and\ \bibinfo {author}
  {\bibfnamefont {C.}~\bibnamefont {Yuan}},\ }\href@noop {} {\  (\bibinfo
  {year} {2026}{\natexlab{a}})},\ \Eprint {http://arxiv.org/abs/2606.21439}
  {arXiv:2606.21439 [gr-qc]} \BibitemShut {NoStop}%
\bibitem [{\citenamefont {Xu}\ \emph {et~al.}(2026{\natexlab{b}})\citenamefont
  {Xu}, \citenamefont {Brito}, \citenamefont {Della~Monica}, \citenamefont
  {Vicente},\ and\ \citenamefont {Yuan}}]{Xu:2026cky}%
  \BibitemOpen
  \bibfield  {author} {\bibinfo {author} {\bibfnamefont {Q.-X.}\ \bibnamefont
  {Xu}}, \bibinfo {author} {\bibfnamefont {R.}~\bibnamefont {Brito}}, \bibinfo
  {author} {\bibfnamefont {R.}~\bibnamefont {Della~Monica}}, \bibinfo {author}
  {\bibfnamefont {R.}~\bibnamefont {Vicente}}, \ and\ \bibinfo {author}
  {\bibfnamefont {C.}~\bibnamefont {Yuan}},\ }\href@noop {} {\  (\bibinfo
  {year} {2026}{\natexlab{b}})},\ \Eprint {http://arxiv.org/abs/2605.03756}
  {arXiv:2605.03756 [gr-qc]} \BibitemShut {NoStop}%
\bibitem [{\citenamefont {Brito}\ and\ \citenamefont
  {Shah}(2023)}]{Brito:2023pyl}%
  \BibitemOpen
  \bibfield  {author} {\bibinfo {author} {\bibfnamefont {R.}~\bibnamefont
  {Brito}}\ and\ \bibinfo {author} {\bibfnamefont {S.}~\bibnamefont {Shah}},\
  }\href {\doibase 10.1103/PhysRevD.108.084019} {\bibfield  {journal} {\bibinfo
   {journal} {Phys. Rev. D}\ }\textbf {\bibinfo {volume} {108}},\ \bibinfo
  {pages} {084019} (\bibinfo {year} {2023})},\ \bibinfo {note} {[Erratum:
  Phys.Rev.D 110, 109902 (2024)]},\ \Eprint {http://arxiv.org/abs/2307.16093}
  {arXiv:2307.16093 [gr-qc]} \BibitemShut {NoStop}%
\bibitem [{\citenamefont {Duque}\ \emph {et~al.}(2024)\citenamefont {Duque},
  \citenamefont {Macedo}, \citenamefont {Vicente},\ and\ \citenamefont
  {Cardoso}}]{Duque:2023seg}%
  \BibitemOpen
  \bibfield  {author} {\bibinfo {author} {\bibfnamefont {F.}~\bibnamefont
  {Duque}}, \bibinfo {author} {\bibfnamefont {C.~F.~B.}\ \bibnamefont
  {Macedo}}, \bibinfo {author} {\bibfnamefont {R.}~\bibnamefont {Vicente}}, \
  and\ \bibinfo {author} {\bibfnamefont {V.}~\bibnamefont {Cardoso}},\ }\href
  {\doibase 10.1103/PhysRevLett.133.121404} {\bibfield  {journal} {\bibinfo
  {journal} {Phys. Rev. Lett.}\ }\textbf {\bibinfo {volume} {133}},\ \bibinfo
  {pages} {121404} (\bibinfo {year} {2024})},\ \Eprint
  {http://arxiv.org/abs/2312.06767} {arXiv:2312.06767 [gr-qc]} \BibitemShut
  {NoStop}%
\bibitem [{\citenamefont {Li}\ \emph {et~al.}(2025)\citenamefont {Li},
  \citenamefont {Weller}, \citenamefont {Bourg}, \citenamefont {LaHaye},
  \citenamefont {Yunes},\ and\ \citenamefont {Yang}}]{Li:2025ffh}%
  \BibitemOpen
  \bibfield  {author} {\bibinfo {author} {\bibfnamefont {D.}~\bibnamefont
  {Li}}, \bibinfo {author} {\bibfnamefont {C.}~\bibnamefont {Weller}}, \bibinfo
  {author} {\bibfnamefont {P.}~\bibnamefont {Bourg}}, \bibinfo {author}
  {\bibfnamefont {M.}~\bibnamefont {LaHaye}}, \bibinfo {author} {\bibfnamefont
  {N.}~\bibnamefont {Yunes}}, \ and\ \bibinfo {author} {\bibfnamefont
  {H.}~\bibnamefont {Yang}},\ }\href {\doibase 10.1103/7l9s-g21j} {\bibfield
  {journal} {\bibinfo  {journal} {Phys. Rev. D}\ }\textbf {\bibinfo {volume}
  {112}},\ \bibinfo {pages} {084057} (\bibinfo {year} {2025})},\ \Eprint
  {http://arxiv.org/abs/2507.02045} {arXiv:2507.02045 [gr-qc]} \BibitemShut
  {NoStop}%
\bibitem [{\citenamefont {Dyson}\ \emph {et~al.}(2025)\citenamefont {Dyson},
  \citenamefont {Spieksma}, \citenamefont {Brito}, \citenamefont {van~de
  Meent},\ and\ \citenamefont {Dolan}}]{Dyson:2025dlj}%
  \BibitemOpen
  \bibfield  {author} {\bibinfo {author} {\bibfnamefont {C.}~\bibnamefont
  {Dyson}}, \bibinfo {author} {\bibfnamefont {T.~F.~M.}\ \bibnamefont
  {Spieksma}}, \bibinfo {author} {\bibfnamefont {R.}~\bibnamefont {Brito}},
  \bibinfo {author} {\bibfnamefont {M.}~\bibnamefont {van~de Meent}}, \ and\
  \bibinfo {author} {\bibfnamefont {S.}~\bibnamefont {Dolan}},\ }\href
  {\doibase 10.1103/PhysRevLett.134.211403} {\bibfield  {journal} {\bibinfo
  {journal} {Phys. Rev. Lett.}\ }\textbf {\bibinfo {volume} {134}},\ \bibinfo
  {pages} {211403} (\bibinfo {year} {2025})},\ \Eprint
  {http://arxiv.org/abs/2501.09806} {arXiv:2501.09806 [gr-qc]} \BibitemShut
  {NoStop}%
\bibitem [{\citenamefont {Datta}\ and\ \citenamefont
  {Maselli}(2026)}]{Datta:2025ruh}%
  \BibitemOpen
  \bibfield  {author} {\bibinfo {author} {\bibfnamefont {S.}~\bibnamefont
  {Datta}}\ and\ \bibinfo {author} {\bibfnamefont {A.}~\bibnamefont
  {Maselli}},\ }\href {\doibase 10.21468/scipostphys.20.3.080} {\bibfield
  {journal} {\bibinfo  {journal} {SciPost Phys.}\ }\textbf {\bibinfo {volume}
  {20}},\ \bibinfo {pages} {080} (\bibinfo {year} {2026})},\ \Eprint
  {http://arxiv.org/abs/2507.04471} {arXiv:2507.04471 [gr-qc]} \BibitemShut
  {NoStop}%
\bibitem [{\citenamefont {Rahman}\ and\ \citenamefont
  {Takahashi}(2026)}]{Rahman:2025mip}%
  \BibitemOpen
  \bibfield  {author} {\bibinfo {author} {\bibfnamefont {M.}~\bibnamefont
  {Rahman}}\ and\ \bibinfo {author} {\bibfnamefont {T.}~\bibnamefont
  {Takahashi}},\ }\href {\doibase 10.1103/76m9-rpzj} {\bibfield  {journal}
  {\bibinfo  {journal} {Phys. Rev. D}\ }\textbf {\bibinfo {volume} {113}},\
  \bibinfo {pages} {044033} (\bibinfo {year} {2026})},\ \Eprint
  {http://arxiv.org/abs/2507.06923} {arXiv:2507.06923 [gr-qc]} \BibitemShut
  {NoStop}%
\bibitem [{\citenamefont {Dyson}\ and\ \citenamefont
  {D'Orazio}(2026)}]{Dyson:2026ddd}%
  \BibitemOpen
  \bibfield  {author} {\bibinfo {author} {\bibfnamefont {C.}~\bibnamefont
  {Dyson}}\ and\ \bibinfo {author} {\bibfnamefont {D.~J.}\ \bibnamefont
  {D'Orazio}},\ }\href@noop {} {\  (\bibinfo {year} {2026})},\ \Eprint
  {http://arxiv.org/abs/2601.19123} {arXiv:2601.19123 [gr-qc]} \BibitemShut
  {NoStop}%
\bibitem [{\citenamefont {Li}\ \emph {et~al.}(2026)\citenamefont {Li},
  \citenamefont {Kuan}, \citenamefont {Weller}, \citenamefont {Yang},
  \citenamefont {Lam}, \citenamefont {Bourg}, \citenamefont {Chung},
  \citenamefont {Chen}, \citenamefont {Yunes},\ and\ \citenamefont
  {Yang}}]{Li:2026foy}%
  \BibitemOpen
  \bibfield  {author} {\bibinfo {author} {\bibfnamefont {D.}~\bibnamefont
  {Li}}, \bibinfo {author} {\bibfnamefont {H.-J.}\ \bibnamefont {Kuan}},
  \bibinfo {author} {\bibfnamefont {C.}~\bibnamefont {Weller}}, \bibinfo
  {author} {\bibfnamefont {C.}~\bibnamefont {Yang}}, \bibinfo {author}
  {\bibfnamefont {K.~K.-H.}\ \bibnamefont {Lam}}, \bibinfo {author}
  {\bibfnamefont {P.}~\bibnamefont {Bourg}}, \bibinfo {author} {\bibfnamefont
  {A.~K.-W.}\ \bibnamefont {Chung}}, \bibinfo {author} {\bibfnamefont
  {Y.}~\bibnamefont {Chen}}, \bibinfo {author} {\bibfnamefont {N.}~\bibnamefont
  {Yunes}}, \ and\ \bibinfo {author} {\bibfnamefont {H.}~\bibnamefont {Yang}},\
  }\href@noop {} {\  (\bibinfo {year} {2026})},\ \Eprint
  {http://arxiv.org/abs/2610.22480} {arXiv:2610.22480 [gr-qc]} \BibitemShut
  {NoStop}%
\bibitem [{\citenamefont {Vicente}\ \emph {et~al.}(2025)\citenamefont
  {Vicente}, \citenamefont {Karydas},\ and\ \citenamefont
  {Bertone}}]{Vicente:2025gsg}%
  \BibitemOpen
  \bibfield  {author} {\bibinfo {author} {\bibfnamefont {R.}~\bibnamefont
  {Vicente}}, \bibinfo {author} {\bibfnamefont {T.~K.}\ \bibnamefont
  {Karydas}}, \ and\ \bibinfo {author} {\bibfnamefont {G.}~\bibnamefont
  {Bertone}},\ }\href {\doibase 10.1103/s4wh-x6c4} {\bibfield  {journal}
  {\bibinfo  {journal} {Phys. Rev. Lett.}\ }\textbf {\bibinfo {volume} {135}},\
  \bibinfo {pages} {211401} (\bibinfo {year} {2025})},\ \Eprint
  {http://arxiv.org/abs/2505.09715} {arXiv:2505.09715 [gr-qc]} \BibitemShut
  {NoStop}%
\bibitem [{\citenamefont {Karydas}\ \emph
  {et~al.}(2026{\natexlab{a}})\citenamefont {Karydas}, \citenamefont
  {Vicente},\ and\ \citenamefont {Bertone}}]{Karydas:2025bkj}%
  \BibitemOpen
  \bibfield  {author} {\bibinfo {author} {\bibfnamefont {T.~K.}\ \bibnamefont
  {Karydas}}, \bibinfo {author} {\bibfnamefont {R.}~\bibnamefont {Vicente}}, \
  and\ \bibinfo {author} {\bibfnamefont {G.}~\bibnamefont {Bertone}},\ }\href
  {\doibase 10.1103/grjg-5k5w} {\bibfield  {journal} {\bibinfo  {journal}
  {Phys. Rev. D}\ }\textbf {\bibinfo {volume} {113}},\ \bibinfo {pages}
  {043057} (\bibinfo {year} {2026}{\natexlab{a}})},\ \Eprint
  {http://arxiv.org/abs/2510.13604} {arXiv:2510.13604 [gr-qc]} \BibitemShut
  {NoStop}%
\bibitem [{\citenamefont {Karydas}\ and\ \citenamefont
  {Bertone}(2026)}]{Karydas:2025gqp}%
  \BibitemOpen
  \bibfield  {author} {\bibinfo {author} {\bibfnamefont {T.~K.}\ \bibnamefont
  {Karydas}}\ and\ \bibinfo {author} {\bibfnamefont {G.}~\bibnamefont
  {Bertone}},\ }\href {\doibase 10.1103/vv8y-m96j} {\bibfield  {journal}
  {\bibinfo  {journal} {Phys. Rev. D}\ }\textbf {\bibinfo {volume} {114}},\
  \bibinfo {pages} {023054} (\bibinfo {year} {2026})},\ \Eprint
  {http://arxiv.org/abs/2512.10855} {arXiv:2512.10855 [gr-qc]} \BibitemShut
  {NoStop}%
\bibitem [{\citenamefont {Barack}(2009)}]{Barack:2009ux}%
  \BibitemOpen
  \bibfield  {author} {\bibinfo {author} {\bibfnamefont {L.}~\bibnamefont
  {Barack}},\ }\href {\doibase 10.1088/0264-9381/26/21/213001} {\bibfield
  {journal} {\bibinfo  {journal} {Class. Quantum Grav.}\ }\textbf {\bibinfo
  {volume} {26}},\ \bibinfo {pages} {213001} (\bibinfo {year} {2009})},\
  \Eprint {http://arxiv.org/abs/0908.1664} {arXiv:0908.1664 [gr-qc]}
  \BibitemShut {NoStop}%
\bibitem [{\citenamefont {Barack}\ and\ \citenamefont
  {Pound}(2019)}]{Barack:2018yvs}%
  \BibitemOpen
  \bibfield  {author} {\bibinfo {author} {\bibfnamefont {L.}~\bibnamefont
  {Barack}}\ and\ \bibinfo {author} {\bibfnamefont {A.}~\bibnamefont {Pound}},\
  }\href {\doibase 10.1088/1361-6633/aae552} {\bibfield  {journal} {\bibinfo
  {journal} {Rep. Prog. Phys.}\ }\textbf {\bibinfo {volume} {82}},\ \bibinfo
  {pages} {016904} (\bibinfo {year} {2019})},\ \Eprint
  {http://arxiv.org/abs/1805.10385} {arXiv:1805.10385 [gr-qc]} \BibitemShut
  {NoStop}%
\bibitem [{\citenamefont {Pound}\ and\ \citenamefont
  {Wardell}(2021)}]{Pound:2021qin}%
  \BibitemOpen
  \bibfield  {author} {\bibinfo {author} {\bibfnamefont {A.}~\bibnamefont
  {Pound}}\ and\ \bibinfo {author} {\bibfnamefont {B.}~\bibnamefont
  {Wardell}},\ }\href {\doibase 10.1007/978-981-15-4702-7\_38-1} {\  (\bibinfo
  {year} {2021}),\ 10.1007/978-981-15-4702-7\_38-1},\ \Eprint
  {http://arxiv.org/abs/2101.04592} {arXiv:2101.04592 [gr-qc]} \BibitemShut
  {NoStop}%
\bibitem [{\citenamefont {Poisson}\ \emph {et~al.}(2011)\citenamefont
  {Poisson}, \citenamefont {Pound},\ and\ \citenamefont
  {Vega}}]{Poisson:2011nh}%
  \BibitemOpen
  \bibfield  {author} {\bibinfo {author} {\bibfnamefont {E.}~\bibnamefont
  {Poisson}}, \bibinfo {author} {\bibfnamefont {A.}~\bibnamefont {Pound}}, \
  and\ \bibinfo {author} {\bibfnamefont {I.}~\bibnamefont {Vega}},\ }\href
  {\doibase 10.12942/lrr-2011-7} {\bibfield  {journal} {\bibinfo  {journal}
  {Living Rev. Relativ.}\ }\textbf {\bibinfo {volume} {14}},\ \bibinfo {pages}
  {7} (\bibinfo {year} {2011})},\ \Eprint {http://arxiv.org/abs/1102.0529}
  {arXiv:1102.0529 [gr-qc]} \BibitemShut {NoStop}%
\bibitem [{\citenamefont {Kavanagh}\ \emph {et~al.}(2025)\citenamefont
  {Kavanagh}, \citenamefont {Karydas}, \citenamefont {Bertone}, \citenamefont
  {Di~Cintio},\ and\ \citenamefont {Pasquato}}]{Kavanagh:2024lgq}%
  \BibitemOpen
  \bibfield  {author} {\bibinfo {author} {\bibfnamefont {B.~J.}\ \bibnamefont
  {Kavanagh}}, \bibinfo {author} {\bibfnamefont {T.~K.}\ \bibnamefont
  {Karydas}}, \bibinfo {author} {\bibfnamefont {G.}~\bibnamefont {Bertone}},
  \bibinfo {author} {\bibfnamefont {P.}~\bibnamefont {Di~Cintio}}, \ and\
  \bibinfo {author} {\bibfnamefont {M.}~\bibnamefont {Pasquato}},\ }\href
  {\doibase 10.1103/PhysRevD.111.063071} {\bibfield  {journal} {\bibinfo
  {journal} {Phys. Rev. D}\ }\textbf {\bibinfo {volume} {111}},\ \bibinfo
  {pages} {063071} (\bibinfo {year} {2025})},\ \Eprint
  {http://arxiv.org/abs/2402.13762} {arXiv:2402.13762 [gr-qc]} \BibitemShut
  {NoStop}%
\bibitem [{\citenamefont {Kavanagh}\ \emph {et~al.}(2020)\citenamefont
  {Kavanagh}, \citenamefont {Nichols}, \citenamefont {Bertone},\ and\
  \citenamefont {Gaggero}}]{Kavanagh:2020cfn}%
  \BibitemOpen
  \bibfield  {author} {\bibinfo {author} {\bibfnamefont {B.~J.}\ \bibnamefont
  {Kavanagh}}, \bibinfo {author} {\bibfnamefont {D.~A.}\ \bibnamefont
  {Nichols}}, \bibinfo {author} {\bibfnamefont {G.}~\bibnamefont {Bertone}}, \
  and\ \bibinfo {author} {\bibfnamefont {D.}~\bibnamefont {Gaggero}},\ }\href
  {\doibase 10.1103/PhysRevD.102.083006} {\bibfield  {journal} {\bibinfo
  {journal} {Phys. Rev. D}\ }\textbf {\bibinfo {volume} {102}},\ \bibinfo
  {pages} {083006} (\bibinfo {year} {2020})},\ \Eprint
  {http://arxiv.org/abs/2002.12811} {arXiv:2002.12811 [gr-qc]} \BibitemShut
  {NoStop}%
\bibitem [{\citenamefont {Coogan}\ \emph {et~al.}(2022)\citenamefont {Coogan},
  \citenamefont {Bertone}, \citenamefont {Gaggero}, \citenamefont {Kavanagh},\
  and\ \citenamefont {Nichols}}]{Coogan:2021uqv}%
  \BibitemOpen
  \bibfield  {author} {\bibinfo {author} {\bibfnamefont {A.}~\bibnamefont
  {Coogan}}, \bibinfo {author} {\bibfnamefont {G.}~\bibnamefont {Bertone}},
  \bibinfo {author} {\bibfnamefont {D.}~\bibnamefont {Gaggero}}, \bibinfo
  {author} {\bibfnamefont {B.~J.}\ \bibnamefont {Kavanagh}}, \ and\ \bibinfo
  {author} {\bibfnamefont {D.~A.}\ \bibnamefont {Nichols}},\ }\href {\doibase
  10.1103/PhysRevD.105.043009} {\bibfield  {journal} {\bibinfo  {journal}
  {Phys. Rev. D}\ }\textbf {\bibinfo {volume} {105}},\ \bibinfo {pages}
  {043009} (\bibinfo {year} {2022})},\ \Eprint
  {http://arxiv.org/abs/2108.04154} {arXiv:2108.04154 [gr-qc]} \BibitemShut
  {NoStop}%
\bibitem [{\citenamefont {Nichols}\ \emph {et~al.}(2023)\citenamefont
  {Nichols}, \citenamefont {Wade},\ and\ \citenamefont
  {Grant}}]{Nichols:2023ufs}%
  \BibitemOpen
  \bibfield  {author} {\bibinfo {author} {\bibfnamefont {D.~A.}\ \bibnamefont
  {Nichols}}, \bibinfo {author} {\bibfnamefont {B.~A.}\ \bibnamefont {Wade}}, \
  and\ \bibinfo {author} {\bibfnamefont {A.~M.}\ \bibnamefont {Grant}},\ }\href
  {\doibase 10.1103/PhysRevD.108.124062} {\bibfield  {journal} {\bibinfo
  {journal} {Phys. Rev. D}\ }\textbf {\bibinfo {volume} {108}},\ \bibinfo
  {pages} {124062} (\bibinfo {year} {2023})},\ \Eprint
  {http://arxiv.org/abs/2309.06498} {arXiv:2309.06498 [gr-qc]} \BibitemShut
  {NoStop}%
\bibitem [{\citenamefont {Wade}\ and\ \citenamefont
  {Nichols}(2026)}]{Wade:2025rkk}%
  \BibitemOpen
  \bibfield  {author} {\bibinfo {author} {\bibfnamefont {B.~A.}\ \bibnamefont
  {Wade}}\ and\ \bibinfo {author} {\bibfnamefont {D.~A.}\ \bibnamefont
  {Nichols}},\ }\href {\doibase 10.1103/5s63-kgx2} {\bibfield  {journal}
  {\bibinfo  {journal} {Phys. Rev. D}\ }\textbf {\bibinfo {volume} {113}},\
  \bibinfo {pages} {044014} (\bibinfo {year} {2026})},\ \Eprint
  {http://arxiv.org/abs/2508.21132} {arXiv:2508.21132 [gr-qc]} \BibitemShut
  {NoStop}%
\bibitem [{\citenamefont {Wade}\ \emph {et~al.}(2026)\citenamefont {Wade},
  \citenamefont {Heeck},\ and\ \citenamefont {Nichols}}]{Wade:2026xvm}%
  \BibitemOpen
  \bibfield  {author} {\bibinfo {author} {\bibfnamefont {B.~A.}\ \bibnamefont
  {Wade}}, \bibinfo {author} {\bibfnamefont {J.}~\bibnamefont {Heeck}}, \ and\
  \bibinfo {author} {\bibfnamefont {D.~A.}\ \bibnamefont {Nichols}},\
  }\href@noop {} {\  (\bibinfo {year} {2026})},\ \Eprint
  {http://arxiv.org/abs/2608.26274} {arXiv:2608.26274 [gr-qc]} \BibitemShut
  {NoStop}%
\bibitem [{\citenamefont {{Chandrasekhar}}(1943)}]{1943ApJ....97..255C}%
  \BibitemOpen
  \bibfield  {author} {\bibinfo {author} {\bibfnamefont {S.}~\bibnamefont
  {{Chandrasekhar}}},\ }\href {\doibase 10.1086/144517} {\bibfield  {journal}
  {\bibinfo  {journal} {Astrophys. J.}\ }\textbf {\bibinfo {volume} {97}},\
  \bibinfo {pages} {255} (\bibinfo {year} {1943})}\BibitemShut {NoStop}%
\bibitem [{\citenamefont {Barausse}(2007)}]{Barausse:2007ph}%
  \BibitemOpen
  \bibfield  {author} {\bibinfo {author} {\bibfnamefont {E.}~\bibnamefont
  {Barausse}},\ }\href {\doibase 10.1111/j.1365-2966.2007.12408.x} {\bibfield
  {journal} {\bibinfo  {journal} {Mon. Not. R. Astron. Soc.}\ }\textbf
  {\bibinfo {volume} {382}},\ \bibinfo {pages} {826} (\bibinfo {year}
  {2007})},\ \Eprint {http://arxiv.org/abs/0709.0211} {arXiv:0709.0211
  [astro-ph]} \BibitemShut {NoStop}%
\bibitem [{\citenamefont {Katz}\ \emph {et~al.}(2019)\citenamefont {Katz},
  \citenamefont {Kurkela},\ and\ \citenamefont {Soloviev}}]{Katz:2019rgf}%
  \BibitemOpen
  \bibfield  {author} {\bibinfo {author} {\bibfnamefont {A.}~\bibnamefont
  {Katz}}, \bibinfo {author} {\bibfnamefont {A.}~\bibnamefont {Kurkela}}, \
  and\ \bibinfo {author} {\bibfnamefont {A.}~\bibnamefont {Soloviev}},\ }\href
  {\doibase 10.1088/1475-7516/2019/08/017} {\bibfield  {journal} {\bibinfo
  {journal} {J. Cosmology Astropart. Phys.}\ }\textbf {\bibinfo {volume}
  {08}},\ \bibinfo {pages} {017} (\bibinfo {year} {2019})},\ \Eprint
  {http://arxiv.org/abs/1906.05341} {arXiv:1906.05341 [hep-ph]} \BibitemShut
  {NoStop}%
\bibitem [{\citenamefont {Traykova}\ \emph {et~al.}(2023)\citenamefont
  {Traykova}, \citenamefont {Vicente}, \citenamefont {Clough}, \citenamefont
  {Helfer}, \citenamefont {Berti}, \citenamefont {Ferreira},\ and\
  \citenamefont {Hui}}]{Traykova:2023qyv}%
  \BibitemOpen
  \bibfield  {author} {\bibinfo {author} {\bibfnamefont {D.}~\bibnamefont
  {Traykova}}, \bibinfo {author} {\bibfnamefont {R.}~\bibnamefont {Vicente}},
  \bibinfo {author} {\bibfnamefont {K.}~\bibnamefont {Clough}}, \bibinfo
  {author} {\bibfnamefont {T.}~\bibnamefont {Helfer}}, \bibinfo {author}
  {\bibfnamefont {E.}~\bibnamefont {Berti}}, \bibinfo {author} {\bibfnamefont
  {P.~G.}\ \bibnamefont {Ferreira}}, \ and\ \bibinfo {author} {\bibfnamefont
  {L.}~\bibnamefont {Hui}},\ }\href {\doibase 10.1103/PhysRevD.108.L121502}
  {\bibfield  {journal} {\bibinfo  {journal} {Phys. Rev. D}\ }\textbf {\bibinfo
  {volume} {108}},\ \bibinfo {pages} {L121502} (\bibinfo {year} {2023})},\
  \Eprint {http://arxiv.org/abs/2305.10492} {arXiv:2305.10492 [gr-qc]}
  \BibitemShut {NoStop}%
\bibitem [{\citenamefont {Annulli}\ \emph {et~al.}(2020)\citenamefont
  {Annulli}, \citenamefont {Cardoso},\ and\ \citenamefont
  {Vicente}}]{Annulli:2020ilw}%
  \BibitemOpen
  \bibfield  {author} {\bibinfo {author} {\bibfnamefont {L.}~\bibnamefont
  {Annulli}}, \bibinfo {author} {\bibfnamefont {V.}~\bibnamefont {Cardoso}}, \
  and\ \bibinfo {author} {\bibfnamefont {R.}~\bibnamefont {Vicente}},\ }\href
  {\doibase 10.1016/j.physletb.2020.135944} {\bibfield  {journal} {\bibinfo
  {journal} {Phys. Lett. B}\ }\textbf {\bibinfo {volume} {811}},\ \bibinfo
  {pages} {135944} (\bibinfo {year} {2020})},\ \Eprint
  {http://arxiv.org/abs/2007.03700} {arXiv:2007.03700 [astro-ph.HE]}
  \BibitemShut {NoStop}%
\bibitem [{\citenamefont {Vicente}\ and\ \citenamefont
  {Cardoso}(2022)}]{Vicente:2022ivh}%
  \BibitemOpen
  \bibfield  {author} {\bibinfo {author} {\bibfnamefont {R.}~\bibnamefont
  {Vicente}}\ and\ \bibinfo {author} {\bibfnamefont {V.}~\bibnamefont
  {Cardoso}},\ }\href {\doibase 10.1103/PhysRevD.105.083008} {\bibfield
  {journal} {\bibinfo  {journal} {Phys. Rev. D}\ }\textbf {\bibinfo {volume}
  {105}},\ \bibinfo {pages} {083008} (\bibinfo {year} {2022})},\ \Eprint
  {http://arxiv.org/abs/2201.08854} {arXiv:2201.08854 [gr-qc]} \BibitemShut
  {NoStop}%
\bibitem [{\citenamefont {Mitra}\ and\ \citenamefont
  {Wilson-Gerow}(2026)}]{Mitra:2026gar}%
  \BibitemOpen
  \bibfield  {author} {\bibinfo {author} {\bibfnamefont {S.}~\bibnamefont
  {Mitra}}\ and\ \bibinfo {author} {\bibfnamefont {J.}~\bibnamefont
  {Wilson-Gerow}},\ }\href@noop {} {\  (\bibinfo {year} {2026})},\ \Eprint
  {http://arxiv.org/abs/2609.28641} {arXiv:2609.28641 [gr-qc]} \BibitemShut
  {NoStop}%
\bibitem [{\citenamefont {Speeney}\ \emph {et~al.}(2022)\citenamefont
  {Speeney}, \citenamefont {Antonelli}, \citenamefont {Baibhav},\ and\
  \citenamefont {Berti}}]{Speeney:2022ryg}%
  \BibitemOpen
  \bibfield  {author} {\bibinfo {author} {\bibfnamefont {N.}~\bibnamefont
  {Speeney}}, \bibinfo {author} {\bibfnamefont {A.}~\bibnamefont {Antonelli}},
  \bibinfo {author} {\bibfnamefont {V.}~\bibnamefont {Baibhav}}, \ and\
  \bibinfo {author} {\bibfnamefont {E.}~\bibnamefont {Berti}},\ }\href
  {\doibase 10.1103/PhysRevD.106.044027} {\bibfield  {journal} {\bibinfo
  {journal} {Phys. Rev. D}\ }\textbf {\bibinfo {volume} {106}},\ \bibinfo
  {pages} {044027} (\bibinfo {year} {2022})},\ \Eprint
  {http://arxiv.org/abs/2204.12508} {arXiv:2204.12508 [gr-qc]} \BibitemShut
  {NoStop}%
\bibitem [{\citenamefont {Mitra}\ \emph {et~al.}(2024)\citenamefont {Mitra},
  \citenamefont {Chakraborty}, \citenamefont {Vicente},\ and\ \citenamefont
  {Feng}}]{Mitra:2023sny}%
  \BibitemOpen
  \bibfield  {author} {\bibinfo {author} {\bibfnamefont {S.}~\bibnamefont
  {Mitra}}, \bibinfo {author} {\bibfnamefont {S.}~\bibnamefont {Chakraborty}},
  \bibinfo {author} {\bibfnamefont {R.}~\bibnamefont {Vicente}}, \ and\
  \bibinfo {author} {\bibfnamefont {J.~C.}\ \bibnamefont {Feng}},\ }\href
  {\doibase 10.1103/PhysRevD.110.084012} {\bibfield  {journal} {\bibinfo
  {journal} {Phys. Rev. D}\ }\textbf {\bibinfo {volume} {110}},\ \bibinfo
  {pages} {084012} (\bibinfo {year} {2024})},\ \Eprint
  {http://arxiv.org/abs/2312.06783} {arXiv:2312.06783 [gr-qc]} \BibitemShut
  {NoStop}%
\bibitem [{\citenamefont {Mitra}\ \emph {et~al.}(2025)\citenamefont {Mitra},
  \citenamefont {Speeney}, \citenamefont {Chakraborty},\ and\ \citenamefont
  {Berti}}]{Mitra:2025tag}%
  \BibitemOpen
  \bibfield  {author} {\bibinfo {author} {\bibfnamefont {S.}~\bibnamefont
  {Mitra}}, \bibinfo {author} {\bibfnamefont {N.}~\bibnamefont {Speeney}},
  \bibinfo {author} {\bibfnamefont {S.}~\bibnamefont {Chakraborty}}, \ and\
  \bibinfo {author} {\bibfnamefont {E.}~\bibnamefont {Berti}},\ }\href
  {\doibase 10.1103/ycl1-kx7d} {\bibfield  {journal} {\bibinfo  {journal}
  {Phys. Rev. D}\ }\textbf {\bibinfo {volume} {112}},\ \bibinfo {pages}
  {044030} (\bibinfo {year} {2025})},\ \Eprint
  {http://arxiv.org/abs/2505.04697} {arXiv:2505.04697 [gr-qc]} \BibitemShut
  {NoStop}%
\bibitem [{\citenamefont {Xiong}\ \emph {et~al.}(2026)\citenamefont {Xiong},
  \citenamefont {Li},\ and\ \citenamefont {Chen}}]{Xiong:2026fqn}%
  \BibitemOpen
  \bibfield  {author} {\bibinfo {author} {\bibfnamefont {W.}~\bibnamefont
  {Xiong}}, \bibinfo {author} {\bibfnamefont {P.-C.}\ \bibnamefont {Li}}, \
  and\ \bibinfo {author} {\bibfnamefont {B.}~\bibnamefont {Chen}},\ }\href@noop
  {} {\  (\bibinfo {year} {2026})},\ \Eprint {http://arxiv.org/abs/2611.11380}
  {arXiv:2611.11380 [gr-qc]} \BibitemShut {NoStop}%
\bibitem [{\citenamefont {Ferrer}\ \emph {et~al.}(2017)\citenamefont {Ferrer},
  \citenamefont {da~Rosa},\ and\ \citenamefont {Will}}]{Ferrer:2017xwm}%
  \BibitemOpen
  \bibfield  {author} {\bibinfo {author} {\bibfnamefont {F.}~\bibnamefont
  {Ferrer}}, \bibinfo {author} {\bibfnamefont {A.~M.}\ \bibnamefont {da~Rosa}},
  \ and\ \bibinfo {author} {\bibfnamefont {C.~M.}\ \bibnamefont {Will}},\
  }\href {\doibase 10.1103/PhysRevD.96.083014} {\bibfield  {journal} {\bibinfo
  {journal} {Phys. Rev. D}\ }\textbf {\bibinfo {volume} {96}},\ \bibinfo
  {pages} {083014} (\bibinfo {year} {2017})},\ \Eprint
  {http://arxiv.org/abs/1707.06302} {arXiv:1707.06302 [astro-ph.CO]}
  \BibitemShut {NoStop}%
\bibitem [{\citenamefont {Kerr}(1963)}]{Kerr:1963ud}%
  \BibitemOpen
  \bibfield  {author} {\bibinfo {author} {\bibfnamefont {R.~P.}\ \bibnamefont
  {Kerr}},\ }\href {\doibase 10.1103/PhysRevLett.11.237} {\bibfield  {journal}
  {\bibinfo  {journal} {Phys. Rev. Lett.}\ }\textbf {\bibinfo {volume} {11}},\
  \bibinfo {pages} {237} (\bibinfo {year} {1963})}\BibitemShut {NoStop}%
\bibitem [{\citenamefont {Wald}(1984)}]{Wald:1984rg}%
  \BibitemOpen
  \bibfield  {author} {\bibinfo {author} {\bibfnamefont {R.~M.}\ \bibnamefont
  {Wald}},\ }\href {\doibase 10.7208/chicago/9780226870373.001.0001} {\emph
  {\bibinfo {title} {{General Relativity}}}}\ (\bibinfo  {publisher} {Chicago
  Univ. Pr.},\ \bibinfo {address} {Chicago, USA},\ \bibinfo {year}
  {1984})\BibitemShut {NoStop}%
\bibitem [{\citenamefont {Datta}(2026)}]{Datta:2026eqj}%
  \BibitemOpen
  \bibfield  {author} {\bibinfo {author} {\bibfnamefont {S.}~\bibnamefont
  {Datta}},\ }\href@noop {} {\  (\bibinfo {year} {2026})},\ \Eprint
  {http://arxiv.org/abs/2608.24800} {arXiv:2608.24800 [gr-qc]} \BibitemShut
  {NoStop}%
\bibitem [{\citenamefont {Wilson-Gerow}(2026)}]{Wilson-Gerow:2026plu}%
  \BibitemOpen
  \bibfield  {author} {\bibinfo {author} {\bibfnamefont {J.}~\bibnamefont
  {Wilson-Gerow}},\ }\href@noop {} {\  (\bibinfo {year} {2026})},\ \Eprint
  {http://arxiv.org/abs/2609.23192} {arXiv:2609.23192 [gr-qc]} \BibitemShut
  {NoStop}%
\bibitem [{\citenamefont {Modrekiladze}\ \emph {et~al.}(2025)\citenamefont
  {Modrekiladze}, \citenamefont {Rothstein},\ and\ \citenamefont
  {Wilson-Gerow}}]{Modrekiladze:2024htc}%
  \BibitemOpen
  \bibfield  {author} {\bibinfo {author} {\bibfnamefont {B.}~\bibnamefont
  {Modrekiladze}}, \bibinfo {author} {\bibfnamefont {I.~Z.}\ \bibnamefont
  {Rothstein}}, \ and\ \bibinfo {author} {\bibfnamefont {J.}~\bibnamefont
  {Wilson-Gerow}},\ }\href {\doibase 10.1088/1475-7516/2025/04/014} {\bibfield
  {journal} {\bibinfo  {journal} {J. Cosmology Astropart. Phys.}\ }\textbf
  {\bibinfo {volume} {04}},\ \bibinfo {pages} {014} (\bibinfo {year} {2025})},\
  \Eprint {http://arxiv.org/abs/2412.06747} {arXiv:2412.06747 [gr-qc]}
  \BibitemShut {NoStop}%
\bibitem [{\citenamefont {Modrekiladze}\ \emph {et~al.}(2026)\citenamefont
  {Modrekiladze}, \citenamefont {Rothstein},\ and\ \citenamefont
  {Wilson-Gerow}}]{Modrekiladze:2026twz}%
  \BibitemOpen
  \bibfield  {author} {\bibinfo {author} {\bibfnamefont {B.}~\bibnamefont
  {Modrekiladze}}, \bibinfo {author} {\bibfnamefont {I.~Z.}\ \bibnamefont
  {Rothstein}}, \ and\ \bibinfo {author} {\bibfnamefont {J.}~\bibnamefont
  {Wilson-Gerow}},\ }\href@noop {} {\  (\bibinfo {year} {2026})},\ \Eprint
  {http://arxiv.org/abs/2609.01644} {arXiv:2609.01644 [gr-qc]} \BibitemShut
  {NoStop}%
\bibitem [{\citenamefont {Hughes}(2000)}]{Hughes:1999bq}%
  \BibitemOpen
  \bibfield  {author} {\bibinfo {author} {\bibfnamefont {S.~A.}\ \bibnamefont
  {Hughes}},\ }\href {\doibase 10.1103/PhysRevD.65.069902} {\bibfield
  {journal} {\bibinfo  {journal} {Phys. Rev. D}\ }\textbf {\bibinfo {volume}
  {61}},\ \bibinfo {pages} {084004} (\bibinfo {year} {2000})},\ \bibinfo {note}
  {[Erratum: Phys.Rev.D 63, 049902 (2001), Erratum: Phys.Rev.D 65, 069902
  (2002), Erratum: Phys.Rev.D 67, 089901 (2003), Erratum: Phys.Rev.D 78, 109902
  (2008), Erratum: Phys.Rev.D 90, 109904 (2014)]},\ \Eprint
  {http://arxiv.org/abs/gr-qc/9910091} {arXiv:gr-qc/9910091} \BibitemShut
  {NoStop}%
\bibitem [{\citenamefont {Ostriker}(1999)}]{Ostriker:1998fa}%
  \BibitemOpen
  \bibfield  {author} {\bibinfo {author} {\bibfnamefont {E.~C.}\ \bibnamefont
  {Ostriker}},\ }\href {\doibase 10.1086/306858} {\bibfield  {journal}
  {\bibinfo  {journal} {Astrophys. J.}\ }\textbf {\bibinfo {volume} {513}},\
  \bibinfo {pages} {252} (\bibinfo {year} {1999})},\ \Eprint
  {http://arxiv.org/abs/astro-ph/9810324} {arXiv:astro-ph/9810324} \BibitemShut
  {NoStop}%
\bibitem [{\citenamefont {Kim}\ and\ \citenamefont {Kim}(2007)}]{Kim:2007zb}%
  \BibitemOpen
  \bibfield  {author} {\bibinfo {author} {\bibfnamefont {H.}~\bibnamefont
  {Kim}}\ and\ \bibinfo {author} {\bibfnamefont {W.-T.}\ \bibnamefont {Kim}},\
  }\href {\doibase 10.1086/519302} {\bibfield  {journal} {\bibinfo  {journal}
  {Astrophys. J.}\ }\textbf {\bibinfo {volume} {665}},\ \bibinfo {pages} {432}
  (\bibinfo {year} {2007})},\ \Eprint {http://arxiv.org/abs/0705.0084}
  {arXiv:0705.0084 [astro-ph]} \BibitemShut {NoStop}%
\bibitem [{\citenamefont {Sharpe}\ \emph
  {et~al.}(2026{\natexlab{a}})\citenamefont {Sharpe}, \citenamefont {Ginat},
  \citenamefont {Spieksma}, \citenamefont {Heger},\ and\ \citenamefont
  {Kocsis}}]{Sharpe:2026nqq}%
  \BibitemOpen
  \bibfield  {author} {\bibinfo {author} {\bibfnamefont {C.}~\bibnamefont
  {Sharpe}}, \bibinfo {author} {\bibfnamefont {Y.~B.}\ \bibnamefont {Ginat}},
  \bibinfo {author} {\bibfnamefont {T.~F.~M.}\ \bibnamefont {Spieksma}},
  \bibinfo {author} {\bibfnamefont {A.}~\bibnamefont {Heger}}, \ and\ \bibinfo
  {author} {\bibfnamefont {B.}~\bibnamefont {Kocsis}},\ }\href@noop {} {\
  (\bibinfo {year} {2026}{\natexlab{a}})},\ \Eprint
  {http://arxiv.org/abs/2603.28866} {arXiv:2603.28866 [gr-qc]} \BibitemShut
  {NoStop}%
\bibitem [{\citenamefont {Sharpe}\ \emph
  {et~al.}(2026{\natexlab{b}})\citenamefont {Sharpe}, \citenamefont {Ginat},
  \citenamefont {Spieksma}, \citenamefont {Heger},\ and\ \citenamefont
  {Kocsis}}]{Sharpe:2026kfk}%
  \BibitemOpen
  \bibfield  {author} {\bibinfo {author} {\bibfnamefont {C.}~\bibnamefont
  {Sharpe}}, \bibinfo {author} {\bibfnamefont {Y.~B.}\ \bibnamefont {Ginat}},
  \bibinfo {author} {\bibfnamefont {T.~F.~M.}\ \bibnamefont {Spieksma}},
  \bibinfo {author} {\bibfnamefont {A.}~\bibnamefont {Heger}}, \ and\ \bibinfo
  {author} {\bibfnamefont {B.}~\bibnamefont {Kocsis}},\ }\href@noop {} {\
  (\bibinfo {year} {2026}{\natexlab{b}})},\ \Eprint
  {http://arxiv.org/abs/2609.37234} {arXiv:2609.37234 [gr-qc]} \BibitemShut
  {NoStop}%
\bibitem [{\citenamefont {Karydas}\ \emph
  {et~al.}(2026{\natexlab{b}})\citenamefont {Karydas}, \citenamefont
  {Scarcella}, \citenamefont {Kavanagh},\ and\ \citenamefont
  {Bertone}}]{Karydas:2026ctq}%
  \BibitemOpen
  \bibfield  {author} {\bibinfo {author} {\bibfnamefont {T.~K.}\ \bibnamefont
  {Karydas}}, \bibinfo {author} {\bibfnamefont {F.}~\bibnamefont {Scarcella}},
  \bibinfo {author} {\bibfnamefont {B.~J.}\ \bibnamefont {Kavanagh}}, \ and\
  \bibinfo {author} {\bibfnamefont {G.}~\bibnamefont {Bertone}},\ }\href@noop
  {} {\  (\bibinfo {year} {2026}{\natexlab{b}})},\ \Eprint
  {http://arxiv.org/abs/2606.13761} {arXiv:2606.13761 [astro-ph.GA]}
  \BibitemShut {NoStop}%
\bibitem [{\citenamefont {Lindquist}(1966)}]{Lindquist:1966igj}%
  \BibitemOpen
  \bibfield  {author} {\bibinfo {author} {\bibfnamefont {R.~W.}\ \bibnamefont
  {Lindquist}},\ }\href {\doibase 10.1016/0003-4916(66)90207-7} {\bibfield
  {journal} {\bibinfo  {journal} {Annals Phys.}\ }\textbf {\bibinfo {volume}
  {37}},\ \bibinfo {pages} {487} (\bibinfo {year} {1966})}\BibitemShut
  {NoStop}%
\bibitem [{\citenamefont {Jeans}(1915)}]{10.1093/mnras/76.2.70}%
  \BibitemOpen
  \bibfield  {author} {\bibinfo {author} {\bibfnamefont {J.~H.}\ \bibnamefont
  {Jeans}},\ }\href {\doibase 10.1093/mnras/76.2.70} {\bibfield  {journal}
  {\bibinfo  {journal} {Monthly Notices of the Royal Astronomical Society}\
  }\textbf {\bibinfo {volume} {76}},\ \bibinfo {pages} {70} (\bibinfo {year}
  {1915})},\ \Eprint
  {http://arxiv.org/abs/https://academic.oup.com/mnras/article-pdf/76/2/70/3010618/mnras76-0070.pdf}
  {https://academic.oup.com/mnras/article-pdf/76/2/70/3010618/mnras76-0070.pdf}
  \BibitemShut {NoStop}%
\bibitem [{\citenamefont {Sadeghian}\ \emph {et~al.}(2013)\citenamefont
  {Sadeghian}, \citenamefont {Ferrer},\ and\ \citenamefont
  {Will}}]{Sadeghian:2013laa}%
  \BibitemOpen
  \bibfield  {author} {\bibinfo {author} {\bibfnamefont {L.}~\bibnamefont
  {Sadeghian}}, \bibinfo {author} {\bibfnamefont {F.}~\bibnamefont {Ferrer}}, \
  and\ \bibinfo {author} {\bibfnamefont {C.~M.}\ \bibnamefont {Will}},\ }\href
  {\doibase 10.1103/PhysRevD.88.063522} {\bibfield  {journal} {\bibinfo
  {journal} {Phys. Rev. D}\ }\textbf {\bibinfo {volume} {88}},\ \bibinfo
  {pages} {063522} (\bibinfo {year} {2013})},\ \Eprint
  {http://arxiv.org/abs/1305.2619} {arXiv:1305.2619 [astro-ph.GA]} \BibitemShut
  {NoStop}%
\bibitem [{\citenamefont {Traykova}\ \emph {et~al.}(2021)\citenamefont
  {Traykova}, \citenamefont {Clough}, \citenamefont {Helfer}, \citenamefont
  {Berti}, \citenamefont {Ferreira},\ and\ \citenamefont
  {Hui}}]{Traykova:2021dua}%
  \BibitemOpen
  \bibfield  {author} {\bibinfo {author} {\bibfnamefont {D.}~\bibnamefont
  {Traykova}}, \bibinfo {author} {\bibfnamefont {K.}~\bibnamefont {Clough}},
  \bibinfo {author} {\bibfnamefont {T.}~\bibnamefont {Helfer}}, \bibinfo
  {author} {\bibfnamefont {E.}~\bibnamefont {Berti}}, \bibinfo {author}
  {\bibfnamefont {P.~G.}\ \bibnamefont {Ferreira}}, \ and\ \bibinfo {author}
  {\bibfnamefont {L.}~\bibnamefont {Hui}},\ }\href {\doibase
  10.1103/PhysRevD.104.103014} {\bibfield  {journal} {\bibinfo  {journal}
  {Phys. Rev. D}\ }\textbf {\bibinfo {volume} {104}},\ \bibinfo {pages}
  {103014} (\bibinfo {year} {2021})},\ \Eprint
  {http://arxiv.org/abs/2106.08280} {arXiv:2106.08280 [gr-qc]} \BibitemShut
  {NoStop}%
\bibitem [{\citenamefont {Chandrasekhar}(1985)}]{Chandrasekhar:1985kt}%
  \BibitemOpen
  \bibfield  {author} {\bibinfo {author} {\bibfnamefont {S.}~\bibnamefont
  {Chandrasekhar}},\ }\href@noop {} {\emph {\bibinfo {title} {{The mathematical
  theory of black holes}}}}\ (\bibinfo {year} {1985})\BibitemShut {NoStop}%
\bibitem [{\citenamefont {Goldreich}\ and\ \citenamefont
  {Tremaine}(1979)}]{Goldreich:1979zz}%
  \BibitemOpen
  \bibfield  {author} {\bibinfo {author} {\bibfnamefont {P.~M.}\ \bibnamefont
  {Goldreich}}\ and\ \bibinfo {author} {\bibfnamefont {S.}~\bibnamefont
  {Tremaine}},\ }\href {\doibase 10.1086/157448} {\bibfield  {journal}
  {\bibinfo  {journal} {Astrophys. J.}\ }\textbf {\bibinfo {volume} {233}},\
  \bibinfo {pages} {857} (\bibinfo {year} {1979})}\BibitemShut {NoStop}%
\bibitem [{\citenamefont {Goldreich}\ and\ \citenamefont
  {Tremaine}(1980)}]{Goldreich:1980wa}%
  \BibitemOpen
  \bibfield  {author} {\bibinfo {author} {\bibfnamefont {P.}~\bibnamefont
  {Goldreich}}\ and\ \bibinfo {author} {\bibfnamefont {S.}~\bibnamefont
  {Tremaine}},\ }\href {\doibase 10.1086/158356} {\bibfield  {journal}
  {\bibinfo  {journal} {Astrophys. J.}\ }\textbf {\bibinfo {volume} {241}},\
  \bibinfo {pages} {425} (\bibinfo {year} {1980})}\BibitemShut {NoStop}%
\bibitem [{\citenamefont {Ward}(1997)}]{Ward:1997di}%
  \BibitemOpen
  \bibfield  {author} {\bibinfo {author} {\bibfnamefont {W.~R.}\ \bibnamefont
  {Ward}},\ }\href {\doibase 10.1006/icar.1996.5647} {\bibfield  {journal}
  {\bibinfo  {journal} {Icarus}\ }\textbf {\bibinfo {volume} {126}},\ \bibinfo
  {pages} {261} (\bibinfo {year} {1997})}\BibitemShut {NoStop}%
\bibitem [{\citenamefont {Gondolo}\ and\ \citenamefont
  {Silk}(1999)}]{Gondolo:1999ef}%
  \BibitemOpen
  \bibfield  {author} {\bibinfo {author} {\bibfnamefont {P.}~\bibnamefont
  {Gondolo}}\ and\ \bibinfo {author} {\bibfnamefont {J.}~\bibnamefont {Silk}},\
  }\href {\doibase 10.1103/PhysRevLett.83.1719} {\bibfield  {journal} {\bibinfo
   {journal} {Phys. Rev. Lett.}\ }\textbf {\bibinfo {volume} {83}},\ \bibinfo
  {pages} {1719} (\bibinfo {year} {1999})},\ \Eprint
  {http://arxiv.org/abs/astro-ph/9906391} {arXiv:astro-ph/9906391} \BibitemShut
  {NoStop}%
\bibitem [{\citenamefont {Lacroix}\ \emph {et~al.}(2018)\citenamefont
  {Lacroix}, \citenamefont {Stref},\ and\ \citenamefont
  {Lavalle}}]{Lacroix:2018qqh}%
  \BibitemOpen
  \bibfield  {author} {\bibinfo {author} {\bibfnamefont {T.}~\bibnamefont
  {Lacroix}}, \bibinfo {author} {\bibfnamefont {M.}~\bibnamefont {Stref}}, \
  and\ \bibinfo {author} {\bibfnamefont {J.}~\bibnamefont {Lavalle}},\ }\href
  {\doibase 10.1088/1475-7516/2018/09/040} {\bibfield  {journal} {\bibinfo
  {journal} {J. Cosmology Astropart. Phys.}\ }\textbf {\bibinfo {volume}
  {09}},\ \bibinfo {pages} {040} (\bibinfo {year} {2018})},\ \Eprint
  {http://arxiv.org/abs/1805.02403} {arXiv:1805.02403 [astro-ph.GA]}
  \BibitemShut {NoStop}%
\bibitem [{\citenamefont {{Young}}(1980)}]{1980ApJ...242.1232Y}%
  \BibitemOpen
  \bibfield  {author} {\bibinfo {author} {\bibfnamefont {P.}~\bibnamefont
  {{Young}}},\ }\href {\doibase 10.1086/158553} {\bibfield  {journal} {\bibinfo
   {journal} {Astrophys. J.}\ }\textbf {\bibinfo {volume} {242}},\ \bibinfo
  {pages} {1232} (\bibinfo {year} {1980})}\BibitemShut {NoStop}%
\bibitem [{\citenamefont {Sadeghian}(2013)}]{Sadeghian:2013bga}%
  \BibitemOpen
  \bibfield  {author} {\bibinfo {author} {\bibfnamefont {L.}~\bibnamefont
  {Sadeghian}},\ }\emph {\bibinfo {title} {{Star Clusters and Dark Matter as
  Probes of the Spacetime Geometry of Massive Black Holes}}},\ \href {\doibase
  10.7936/K7JM27QG} {Ph.D. thesis},\ \bibinfo  {school} {Washington U., St.
  Louis} (\bibinfo {year} {2013}),\ \Eprint {http://arxiv.org/abs/1308.5378}
  {arXiv:1308.5378 [gr-qc]} \BibitemShut {NoStop}%
\bibitem [{\citenamefont {Wang}\ \emph {et~al.}(2026)\citenamefont {Wang},
  \citenamefont {Chadha-Day}, \citenamefont {Clough},\ and\ \citenamefont
  {Maxwell}}]{Wang:2026daj}%
  \BibitemOpen
  \bibfield  {author} {\bibinfo {author} {\bibfnamefont {Z.}~\bibnamefont
  {Wang}}, \bibinfo {author} {\bibfnamefont {F.}~\bibnamefont {Chadha-Day}},
  \bibinfo {author} {\bibfnamefont {K.}~\bibnamefont {Clough}}, \ and\ \bibinfo
  {author} {\bibfnamefont {J.}~\bibnamefont {Maxwell}},\ }\href@noop {} {\
  (\bibinfo {year} {2026})},\ \Eprint {http://arxiv.org/abs/2610.13684}
  {arXiv:2610.13684 [gr-qc]} \BibitemShut {NoStop}%
\bibitem [{\citenamefont {Barnes}\ and\ \citenamefont
  {Efstathiou}(1987)}]{Barnes:1987hu}%
  \BibitemOpen
  \bibfield  {author} {\bibinfo {author} {\bibfnamefont {J.}~\bibnamefont
  {Barnes}}\ and\ \bibinfo {author} {\bibfnamefont {G.}~\bibnamefont
  {Efstathiou}},\ }\href {\doibase 10.1086/165480} {\bibfield  {journal}
  {\bibinfo  {journal} {Astrophys. J.}\ }\textbf {\bibinfo {volume} {319}},\
  \bibinfo {pages} {575} (\bibinfo {year} {1987})}\BibitemShut {NoStop}%
\bibitem [{\citenamefont {Bullock}\ \emph {et~al.}(2001)\citenamefont
  {Bullock}, \citenamefont {Dekel}, \citenamefont {Kolatt}, \citenamefont
  {Kravtsov}, \citenamefont {Klypin}, \citenamefont {Porciani},\ and\
  \citenamefont {Primack}}]{Bullock:2000ry}%
  \BibitemOpen
  \bibfield  {author} {\bibinfo {author} {\bibfnamefont {J.~S.}\ \bibnamefont
  {Bullock}}, \bibinfo {author} {\bibfnamefont {A.}~\bibnamefont {Dekel}},
  \bibinfo {author} {\bibfnamefont {T.~S.}\ \bibnamefont {Kolatt}}, \bibinfo
  {author} {\bibfnamefont {A.~V.}\ \bibnamefont {Kravtsov}}, \bibinfo {author}
  {\bibfnamefont {A.~A.}\ \bibnamefont {Klypin}}, \bibinfo {author}
  {\bibfnamefont {C.}~\bibnamefont {Porciani}}, \ and\ \bibinfo {author}
  {\bibfnamefont {J.~R.}\ \bibnamefont {Primack}},\ }\href {\doibase
  10.1086/321477} {\bibfield  {journal} {\bibinfo  {journal} {Astrophys. J.}\
  }\textbf {\bibinfo {volume} {555}},\ \bibinfo {pages} {240} (\bibinfo {year}
  {2001})},\ \Eprint {http://arxiv.org/abs/astro-ph/0011001}
  {arXiv:astro-ph/0011001} \BibitemShut {NoStop}%
\bibitem [{\citenamefont {Chiba}\ and\ \citenamefont
  {Kataria}(2024)}]{Chiba:2023rqm}%
  \BibitemOpen
  \bibfield  {author} {\bibinfo {author} {\bibfnamefont {R.}~\bibnamefont
  {Chiba}}\ and\ \bibinfo {author} {\bibfnamefont {S.~K.}\ \bibnamefont
  {Kataria}},\ }\href {\doibase 10.1093/mnras/stae288} {\bibfield  {journal}
  {\bibinfo  {journal} {Mon. Not. R. Astron. Soc.}\ }\textbf {\bibinfo {volume}
  {528}},\ \bibinfo {pages} {4115} (\bibinfo {year} {2024})},\ \Eprint
  {http://arxiv.org/abs/2311.07640} {arXiv:2311.07640 [astro-ph.GA]}
  \BibitemShut {NoStop}%
\end{thebibliography}%

\end{document}